\documentclass[sigconf]{acmart}
\usepackage{style/myStyle}

\AtBeginDocument{%
  }

\begin{document}

\title{SILENT: A New Lens on Statistics in Software Timing Side Channels}


\author{Martin Dunsche}
\affiliation{%
  \institution{Ruhr University Bochum}
  \country{Germany}
}
\email{martin.dunsche@rub.de}
\author{Patrick Bastian}
\affiliation{%
  \institution{Ruhr University Bochum}
  \country{Germany}
}
\email{patrick.bastian@rub.de}

\author{Marcel Maehren}
\affiliation{%
  \institution{Ruhr University Bochum}
  \country{Germany}
}
\email{marcel.meahren@rub.de}

\author{Nurullah Erinola}
\affiliation{%
  \institution{Ruhr University Bochum}
  \country{Germany}
}
\email{nurullah.erinola@rub.de}

\author{Robert Merget}
\affiliation{%
  \institution{Technology Innovation Institute}
  \country{United Arab Emirates}
}
\email{robert.merget@tii.ae}

\author{Nicolai Bissantz}
\affiliation{%
  \institution{Ruhr University Bochum}
  \country{Germany}
}
\email{nicolaibernhard.bissantz@rub.de}

\author{Holger Dette}
\affiliation{%
  \institution{Ruhr University Bochum}
  \country{Germany}
}
\email{holger.dette@rub.de}

\author{Jörg Schwenk}
\affiliation{%
  \institution{Ruhr University Bochum}
  \country{Germany}
}
\email{joerg.schwenk@rub.de}
\renewcommand{\shortauthors}{Dunsche et al.}
\settopmatter{printacmref=false}
\renewcommand\footnotetextcopyrightpermission[1]{}
\settopmatter{printfolios=true}

\fancyhead[L]{\textit{\shorttitle}} 
\fancyhead[R]{\textit{Dunsche et al.}}

\renewcommand\footnotetextcopyrightpermission[1]{}


\begin{abstract}

Cryptographic research takes software timing side channels seriously. Approaches to mitigate them include constant-time coding and techniques to enforce such practices. However, recent attacks like Meltdown~\cite{USENIX:LSGPHF18}, Spectre~\cite{SP:KHFGGH19}, and Hertzbleed~\cite{USENIX:WPHSFK22} have challenged our understanding of what it means for code to execute in constant time on modern CPUs. To ensure that assumptions on the underlying hardware are correct and to create a complete feedback loop, developers should also perform \emph{timing measurements} as a final validation step to ensure the absence of exploitable side channels. Unfortunately, as highlighted by a recent study by Jancar et al.~\cite{SP:JFDSSB22}, developers often avoid measurements due to the perceived unreliability of the statistical analysis and its guarantees.

In this work, we combat the view that statistical techniques only provide weak guarantees by introducing a new algorithm for the analysis of timing measurements with strong, formal statistical guarantees, giving developers a reliable analysis tool. Specifically, our algorithm (1) is \textit{non-parametric}, making minimal assumptions about the underlying distribution and thus overcoming limitations of classical tests like the t-test, (2) handles unknown data dependencies in measurements, (3) can estimate in advance how many samples are needed to detect a leak of a given size, and (4) allows the definition of a negligible leak threshold $\Delta$, ensuring that acceptable non-exploitable leaks do not trigger false positives, without compromising statistical soundness. We demonstrate the necessity, effectiveness, and benefits of our approach on both synthetic benchmarks and real-world applications.
\end{abstract}





\settopmatter{printfolios=true}


\maketitle

\section{Introduction}

Since their initial discovery by Paul Kocher \cite{C:Kocher96}, side-channel vulnerabilities have played a crucial role in the security of software implementations, especially in the field of cryptography. 
Software timing-based side-channel vulnerabilities are especially relevant as they can potentially be exploited remotely.  

\paragraph{Constant-Time Implementations}
To avoid timing related vulnerabilities, developers follow \textit{constant time coding practices}, so that the execution time of their code is independent of secret data.. These practices include avoiding branching based on secret data, avoiding secret-based memory access, and avoiding assembly instructions with variable execution times for secret inputs. 
To provide guarantees for such code, other approaches like dynamic instrumentation (e.g., \cite{SP:DanBarRez20, USENIX:WSBS20}) and formal analysis tools (e.g., \cite{CCS:CheFenDil17, CCS:BBCLP14, USENIX:DFKMR13}) have been used, with varying degrees of formal security guarantees.
A study by Jancar et al.~\cite{SP:JFDSSB22} investigates the use of such tools as a means to preventtiming side channels. 
In general, the study found that none of the tool-assisted approaches were widely accepted by developers, with over 60\% of participants reporting no tool usage at all, with many citing poor usability or lack of maintenance as core issues preventing further adoption. 

While these practices are the foundation for modern secure implementations, recent research has shown that our assumptions on how secrets get leaked on modern hardware architectures are continually being subverted \cite{USENIX:WPHSFK22,SP:KHFGGH19,USENIX:LSGPHF18,Ragab2021CrossTalkSD}. Features like speculative execution or CPU frequency scaling are able to introduce variable execution time even when all best practices are being followed. At the same time, it is unknown if the community has identified all potential sources of leakage in a given architecture, leaving an unaddressed risk when analyzing implementations based solely on known best practice. 
\emph{We believe that closing the feedback loop by actually measuring the timing behavior of software implementations after other techniques have been applied will help to reduce this remaining risk significantly.}

\paragraph{Statistical Evaluation of Timing Measurements}
Statistical analyses of side-channel measurements likely pose the fewest challenges in terms of usability and maintenance. For the analysis, the developer has to measure the (unbiased) execution time of the algorithm for different input classes and pass the measurements to a statistical algorithm. Measuring execution time is a trivial task for cryptographic developers, and statistical algorithms do not require a lot of maintenance, as they do not change and do not interact much with compilers, other software libraries, or the operating system. 

However, today's statistical methods to analyze timing measurements have significant drawbacks --  participants from Jancar et al.~\cite{SP:JFDSSB22} reported false positives and limited guarantees as some of the core points against statistical techniques.
To quote Jancar et al.:
\begin{quote}
    These tools are generally easy to install and run, even at scale, and operate on executable code, ruling out the possibility of compiler-induced violations of the constant-time policy. \emph{However, they only provide weak, informal guarantees.}
\end{quote}

In this paper, we address the last point: We extend the set of statistical guarantees that can be given when evaluating timing measurements to mitigate some of the perceived drawbacks.
A first step in this direction was already made by Dunsche et al. \cite{USENIX:DMEMBS24} and Kario \cite{ESORICS:Kario23}, who showed that it is possible to strictly bind the false positive rate when evaluating timing measurements, even when only minor assumptions on the distributions of the measurements are made. 

In this paper, we address different problems to improve the reliability and usability of statistical algorithms for \emph{software} side channels.
\paragraph{Dependent Measurements} 
Timing side channels rely on repeated measurements of the execution time of a protocol, typically gathered by running the protocol many times in succession. To analyze the measurements based on a statistical test, one makes simple assumptions under which the statistical test yields reliable results. A common and crucial assumption in prior work, such as \cite{dudect16,USENIX:DMEMBS24,ESORICS:Kario23}, is the independence between measurements. While this assumption may hold in an optimized setting, it often breaks down in practice due to environmental factors, application-specific behaviors, or many other uncontrolled conditions inside the system under test. 
One simple example is the usage of caches in the operating system and the CPU. Each measurement can alter cache values, affecting subsequent timing measurements.
This brings us to our first research question:
\begin{quote}
\textbf{RQ1:} 
Can statistical tests still yield reliable results when measurements are dependent?
\end{quote}
The question can be answered in the affirmative. 
To demonstrate that addressing data dependency is beneficial, we analyze data from the artifact provided by Dunsche et al.~\cite{USENIX:DMEMBS24}: Timing measurements of the mbedTLS library in the most recent version available at the time, in the context of the Bleichenbacher~\cite{C:Bleichenbacher98} vulnerability.
In this example, the tool from Dunsche et al. reported a side channel, which could not be confirmed by manual code analysis from the researchers and was therefore classified as a false positive. When analyzing the data, we found a strong positive correlation, which likely resulted in the incorrect analysis of their tool (\gls{rtlf}).
By considering dependent measurements, our proposed approach will avoid such misclassifications as we will show in our evaluation.

\paragraph{Negligible Side Channels}

For many applications, the size of a timing leak is a key factor for the consideration of whether it is actually exploitable and whether it is worth investing the time in mitigating it. For example, the impact of micro-architectural leakage is assumed to be small, making such leakage, especially in remote scenarios, likely unexploitable. Many developers, therefore, often exclude this type of leakage from their attacker model. A prominent example of this is the popular cryptographic library OpenSSL, which explicitly excludes side-channel vulnerabilities that are only locally exploitable from their attacker model\footnote{\url{https://openssl-library.org/policies/general/security-policy/}}. At the same time, this mindset is not captured by current statistical approaches. As the number of samples increases, current techniques will eventually detect and report even the smallest leakage as a vulnerability, even when the leakage is too small to actually impact security or stems from a biased measurement setup. This leads us to our second research question:
 
\begin{quote}
\textbf{RQ2:} 
Can a statistical testing framework be designed to ensure that negligible side channels or minor structural biases are not detected as statistically significant differences?
\end{quote}
We answer RQ2 in the affirmative. 
To demonstrate this, we repeat the experiments of Bernstein et al. \cite{bernstein2025kyberslash} using measurement code from their artifact to collect data for their discovered \textit{Kyberslash} leak. 
We then empirically study the impact of varying parameters reflecting different notions of negligible side channels. We do not employ one specific threshold, as Crosby et al.~\cite{crosby09}, for example, considered side channels below 100ns unexploitable, while Hubert Kario argues that side channels as small as \emph{one} clock cycle can be detectable over local area networks~\cite{kario23}. We want to emphasize that smaller timing leaks are generally possible since the CPU frequency may vary, and timing leaks may manifest non-deterministically across samples.
When applying less strict thresholds, as in Crosby's assessment, our algorithm does not flag the leak of the Kyberslash vulnerability - expected to be around 20 clock cycles - as practically significant. On the other hand, following Hubert Kario's considerations for exploitability, our test does indicate a significant leak as expected.
Our statistical algorithm gives developers a flexible tool to apply their own assumptions and considered attacker models enabling sound risk assessments. 

\paragraph{High-level statistical idea}
For both \textbf{RQ1} and \textbf{RQ2}, we developed a resampling-based test procedure that accounts for dependence both between and within the measurements. The test is then embedded in a framework that allows specifying the maximum size $\Delta$ of a potential difference
should not be detected. To be precise, given a reasonable amount of measurements, our statistical test will \textit{never} detect differences
that are smaller than $\Delta$ and even those that are precisely equal to $\Delta$ are only detected at a prescribed rate $\alpha$. In practice, this usually results in \textbf{no false positives} due to structural bias or negligible side channels.

Our new statistical algorithm has additional advantages that will produce more accurate results than previous techniques. In contrast to \gls{rtlf} proposed in Dunsche et al.~\cite{USENIX:DMEMBS24}, our algorithm is able to also handle \textit{discrete} measurements, where only few distinct values are observed. Additionally, our approach can work with arbitrary alternatives, does not require multiple testing, and is non-parametric, which eventually will increase the accuracy of the test.

\paragraph{Restricted measurement environments}
Another question that developers and pentesters commonly face when performing statistical side-channel analysis concerns the number of measurements required to obtain statistically reliable results.
In some scenarios, performing measurements incurs certain costs — such as network latency, long test initialization times, or security restrictions (e.g., intrusion detection systems limiting the number of requests per timeframe) — which can incentivize users to perform fewer measurements than might be necessary to detect a given side channel.
In this work, we provide practitioners with a tool to make accurate and cost-efficient choices for their required sample size by answering the following research question:
\begin{quote}
\textbf{RQ3:} How many measurements do we need to identify a side channel of a given size?
\end{quote}
Our statistical algorithm naturally provides this information by Theorem \ref{thm:bootstrappower}. We will show that this can be used to optimize the measurement process in scenarios where measurements are typically not \textit{free} for the tester.
Specifically, we will show that by first making a few measurements, we can estimate how many measurements we will need to detect a side channel of an assumed size reliably. To demonstrate the reliability of our approach, we revisit the Kyberslash example to predict the amount of measurements required to detect the vulnerability. We then additionally show the practical benefit of our approach for web pentesters, by showing how the tester’s position in the network affects the number of measurements needed to detect a given time leak in an artificially constructed web-based timing side-channel vulnerability.  


\paragraph{Contributions}
\begin{itemize}
    \item We provide a broadly applicable theoretically founded testing framework to analyze timing measurements of software implementations of two input classes that can handle dependent data, considers arbitrary alternatives, has bounded type-1 errors, does not require multiple testing and is non-parametric.
    \item We give a broad range of statistical guarantees which state-of-the-art tools cannot provide. 
    
    \item We implement and provide our approach as an open-source tool for the community.
    \item We showcase the developed methodology on diverse practical applications in both cryptography and web security.  
\end{itemize}

\paragraph{Ethical Considerations}
Since we did only rely on artifacts already published in related work, and on lab experiments with our own web application, there was no need to perform a Repsonsible Disclosure.

\paragraph{Artifacts}
As part of our artifacts, we provide all evaluated real-world measurements, including the specific data sets taken from related work, and results. We further provide the scripts to generate the artificial data for our ground truth study,the setup of the tested applications, and - most importantly - the code used for our statistical evaluations.
\section{Background}\label{sec:background}
In 1996, Paul Kocher introduced the world to timing based side-channel attacks~\cite{C:Kocher96}. In these attacks, the attacker uses the execution time of an algorithm to learn something about its inputs. Since then, many timing attacks have been proposed in the literature targeting various cryptographic systems~\cite{lucky13,luckyMicroseconds15,meyer14,cachebleed17,raccoon21,brumley03,stillPractical11,ecdsaPractical15,Bernstein2005CachetimingAO,rsacrttiming,SP:VanRon20}, but also the operating system~\cite{SP:KHFGGH19,USENIX:LSGPHF18} or the application layer~\cite{bortzweb,hiddencaches,timewilltell,timeprivacy}. To protect against timing attacks, the community created a set of programming paradigms colloquially known as 'constant-time programming'. Under these paradigms, the programmer is not allowed to use branching instructions with secret inputs or perform secret-based memory lookups (to avoid cache-induced timings). Additionally, the programmer should not use assembly instructions with variable execution time and secret inputs. However, recent research has shown that these paradigms are not enough to ensure that code actually runs in secret independent time. Attacks like Hertzbleed~\cite{USENIX:WPHSFK22} or Spectre~\cite{SP:KHFGGH19} subverted our expectations of what is actually happening on the hardware with our code, which questions whether the developed code is actually 'constant time' enough to be secure in actual applications. With potentially further unknown influences on the execution time, measuring the execution time is an intuitive tool to ensure that no significant side channels are present.

\subsection{Statistical Hypothesis Test}
\label{sec:hypothesistest}
Statistical hypothesis tests are used to take a quantitative decision based on data $x:=(x_1,\hdots,x_n)$ on rejection or acceptance of some null hypothesis $H_0$ against an alternative hypothesis $H_1$. Such tests are based on a \textit{summary statistic} $S:=S(x)$, which is a quantitative measure of the plausibility of $H_0$ given some observed data. $S$ is also called the \textit{test statistic} and returns a real-value, such as the sample mean. In the context of side-channel analysis, this is the test that, given the measurements, will decide whether or not a side channel is present.

\paragraph{Type-1 error}
A hypothesis test is constructed such that for a sample $x$, it rejects $H_0$ in favor of $H_1$ if $S(x)$ exceeds some threshold $c$, above which $S$ seems at odds with $H_0$. 
Since the data contains some unpredictable randomness, there is always a risk of a false decision. In statistical terms, rejecting $H_0$ when it holds true is referred to as a type-1 error (\textit{False Positive}), while not rejecting a hypothesis $H_0$, while it is not true, is called a type-2 error (\textit{False Negative}). Here, we define $\alpha$ as the maximal type-1 error rate considered to be acceptable. Specifically, we have $\mathbb{P}(S>c| H_0 \textnormal{ is true})\leq \alpha$. By choosing a high $\alpha$ value, the hypothesis test can make more 'risky' classifications, while a low $\alpha$ value makes the test more conservative.

\paragraph{Statistical Power and Alternatives}
The \textit{power} of a hypothesis test is the probability of rejecting $H_0$ when $H_1$ is true (i.e., the probability of a true positive). As the sample size increases, a well-designed hypothesis test should eventually reject $H_0$ when $H_1$ is indeed true or, in mathematical terms, $\lim_{n\to\infty}\mathbb{P}(S>c| H_1 \,\textnormal{is true})=1$. This is an important characterization for the type-2 error, as it is asymptotically negligible in the sample size $n$. Generally speaking, the statistical power expresses how often the test identifies a correctly positive result and, therefore, always depends on the alternative (the true difference from the null hypothesis) at hand. For larger differences, the power is generally higher than for smaller. 

\paragraph{Multiple Testing}
 In many cases, a null hypothesis is a composition of several sub-hypotheses, and it is rejected when at least one test rejects a sub-hypothesis. Here, one can think of different tests applied to the same data. This yields the classical multiple testing problem (see Noble ~\cite{Noble2009} for an overview), where each separate test can erroneously decide that the null hypothesis is false. For instance, if there are $p$ sub-hypotheses and one performs for each hypothesis a test with a prescribed type-1 error $\alpha$, then the type-1 error rate of the combined test (which rejects if at least one test for a sub-hypothesis rejects) can only be upper bounded by $p\alpha$. Moreover, this upper bound can not be improved without further assumptions. To handle this problem, one can use classical methods like the Bonferroni correction \cite{bonferroni1936}. Slightly more advanced tests are multivariate tests or tests based on functions of the individual tests.
 
\paragraph{Dependent Data}
A classical assumption in statistics are \emph{independent observations}, where each sample is entirely independent of other samples. However, real-world applications, such as timing measurements on real hardware, often introduce dependencies due to the underlying system. Whenever a measurement is done, it influences the system under test. In the context of timing measurements, this can be a subtle modification like the changing of values in the CPU cache, changes in the CPU frequency due to consumed electricity, or a not-so-subtle change in the RAM or hard drive introduced by the application that is tested. Ignoring such dependencies can lead to misleading inferences, such as underestimated variability (noise) and or inflated type-1 errors. To overcome that, it is possible to model this dependency in the statistical analysis to allow for valid conclusions.

\paragraph{Parametric Assumptions}
Many hypothesis tests make assumptions about the expected distributions of the data that will be seen. A simple (and for software timing measurements naive assumption), is that the samples will be normal distributed. If the expected distribution can be expressed by a function with a small set of parameters, we call these distributions parametric. 
A classic example of a test that makes a parametric assumption is the t-test~\cite{Sachs.1984}, where normally distributed data is assumed. A normal distribution can be fully determined by the mean and variance, which makes it parametric. Tests that use parametric assumptions are powerful when the corresponding assumptions are (at least approximately) satisfied but can produce misleading results if these assumptions are violated. In software timing measurements, it is often optimistic to know the distribution in advance. Mistakes that introduce non-constant time behavior can be completely arbitrary and can potentially also overlap, producing distributions that are not parametric anymore.
Whenever we make no parametric assumption, we usually use non-parametric statistics such as quantiles.

\paragraph{Quantiles} The (empirical) $k$-th quantile is a statistical measure that separates the lowest $k$\% of the data set from the remaining $(100-k)$\% of the data. For a formal statistical analysis, the $k$-th quantile of a real-valued random variable $X$ can be defined as the smallest number $q_k$ such that the inequality  $\mathbb{P}(X\leq q_k)\geq k$ holds. In other words, in average $k\%$ of the observations are smaller or equal to $q_k$ and $(1-k)\%$ of the observation are larger. 

\paragraph{Booststrap}\label{subsec:bootstrap}
In many applications, the threshold $c$ (see Type-1 error) required for a statistical test is difficult to obtain by theoretical arguments. However, often $c$ can be estimated through a simple procedure known as \textit{bootstrap}, which is based on a simulation of the test statistic for artificially generated random data with a distribution closely similar to the true but unknown distribution of the data. By repeatedly resampling the data set and imitating the behavior of $S$ under $H_0$, an empirical estimator, say $c_{1-\alpha}^*$,  for $c:= c_{1-\alpha}$ can be obtained. Formally, we consider a rejection rule of the form $S>c$, where we assume that $S$ is resampled exactly $B$ times, say $S_1^*,\hdots, S_B^*$ and denote the ordered values of this procedure by $S_{(1)}^* \leq  \hdots \leq S_{(B)}^*$. This allows us to estimate the unknown threshold $c$ by the empirical threshold $(1-\alpha)$-quantile $c_{1-\alpha}^*:= S_{(\lfloor(1-\alpha)B\rfloor)}$ of this sample.
In summary, bootstrapping allows one to define a \textit{configurable} threshold, and we refer to~\cite{vaart1998} for a more detailed discussion on the subject.

\subsection{Statistical Side-Channel Analysis}
In statistical side-channel analysis, the user repeatedly measures the execution time of an algorithm for two different classes of inputs that \emph{should} have the same execution time. Once the measurements are gathered, the user then applies a statistical test to determine if the measurements follow the same distribution. If they do, the implementation is considered secure (regarding the tested classes of inputs), while if the distributions are different, the implementation is considered insecure. Several approaches have been proposed for such tests. The most important will be discussed in the following: \textbf{\gls{tvla}} is a statistical approach commonly used to search side-channel vulnerabilities in cryptographic primitives~\cite{tvlapublickey,tvlanistlwc,hardwarepktvla}. It was proposed by Goodwill et al. with the Test Vector Leakage Assessment (TVLA) framework~\cite{tvla}. Typically, it is used with Welch's t-test~\cite{Sachs.1984}, the paired t-test~\cite{tvlapairedttest}, or the $\chi^2$-test~\cite{chi2tvla} to obtain a test decision.  \textbf{\gls{dudect}}
~\cite{dudect16} performs a series of Welch's t-tests~\cite{Sachs.1984} on a partitioned data set. \gls{dudect} is designed to measure \textit{and} test repeatedly until the first positive is detected. \textbf{\gls{mona}}
~\cite{mona12} is a testing tool based on Crosby's box test \cite{crosby09}. It iterates over all percentiles and reports a timing leak if it finds a box of size 1\% or bigger. 
\textbf{\gls{tlsfuzzer}}
~\cite{ESORICS:Kario23} performs a collection of well established statistical tests. 
For the final decision, four distinct tests are executed, namely the Wilcoxon signed-rank test~\cite{wilcoxon-signed-rank}, Sign test~\cite[Section 15.3]{math-statistics-book-sign-test}, the paired t-test~\cite{Sachs.1984}, and the Friedman test~\cite{friedman-test}. As illustrated in the statistical background (Multiple Testing), a statistically significant difference is reported whenever at least one test detects one. \textbf{\gls{rtlf}}~\cite{USENIX:DMEMBS24}~
provides a testing procedure that is based on quantiles. Essentially, it considers the deciles as its primal focus and guarantees a bounded type-1 error based on a bootstrap.

\subsection{Relevant Hypothesis Testing}\label{subsec:advantages_relevant_hypothesis}

In hypothesis testing for two distributions $P_X$ and $P_Y$, for instance, representing the execution time of an algorithm, it is common to formulate the hypotheses in terms of exact equality: 
$H_0: P_X = P_Y$ versus $H_1: P_X \neq P_Y$. That is, the null hypothesis assumes exactly identical timing behavior, while the alternative hypothesis assumes a difference. However, in the real world, testing for exact equality is ambitious. As John Tukey~\cite{Tukey1991ThePO} famously noted: \begin{quote}\textit{All we know
about the world teaches us that the effects of A and B are always different—in some decimal
place—for any A and B. Thus asking “Are the effects different?” is foolish}.
\end{quote}
To overcome that simplification, we suggest, among many others in the statistical literature (see Stefan Wellek \cite{wellek2010} for a comprehensive account), to use a {\it relevant hypotheses}. The key difference lies in the construction of the hypotheses. Namely, for a threshold $\Delta>0$ and some distance measure $d$ between the distributions $P_X$ and $P_Y$, we consider the hypotheses
\begin{align}
\label{det1}
H_0: d(P_X,P_Y)\leq \Delta\quad vs.\quad H_1: d(P_X, P_Y)>\Delta~.
\end{align}
In simple terms, rather than asking whether the distributions are exactly identical or not, we assess whether they are sufficiently similar or meaningfully different. 

\paragraph{Practical Significance vs. Statistical Significance}

It is crucial to distinguish between practical significance and statistical significance. With a large sample size, one may detect very tiny differences that might not be relevant for the application and consideration. In the context of side-channel attacks, the size of a side channel indicates if it is likely exploitable or not in a given attacker model. For example, Crosby et al. \cite{crosby09} argue that timing differences of $100 ns$ or larger are likely exploitable in a LAN-setting. Hence, following this assumption, if we aim to assess whether some code or protocol possibly contains an exploitable side channel in a LAN setting, a meaningful test could be designed in a way that it only detects side channels of around $100ns$ or larger. Here, we want to emphasize that the naive approach of rejecting $H_0$ in \eqref{det1}, whenever any chosen test rejects the classical hypotheses \textit{and} the test-statistic exceeds $\Delta$, does not yield a statistical test controlling the type-1 error rate. We illustrate this fact in Appendix \ref{app:example_t_test} for the classical $t$-test.

\paragraph{Controlling Type-2 Errors}

When performing a statistical test, one can always control either the type-1 or the type-2 error (in the overwhelming number of cases, the type-1 error), and given that error, we aim to minimize the other error. In the context of software timing side channels, e.g. Dunsche et al.~\cite{USENIX:DMEMBS24} did this in \gls{rtlf}. They chose a static false positive rate $\alpha$ and then tried to minimize the amount of false negatives, i.e., maximizing the number of correct results. Therefore, a hypothesis test is constructed in a way that the controlled error causes more harm. Obviously, defining what causes more significant harm depends on the application. In the context of side channels, a type-1 error may cause the developer to spend hours trying to debug a vulnerability that is not present. In contrast, a type-2 error risks a vulnerability going undetected, which may leak information to an attacker, potentially inflicting very high financial damage or threat to human lives. Consequently, if it is desirable to control the type-2 error instead of the type-1 error, we would have to switch the hypotheses in \eqref{det1}. However, doing that is often unfeasible in the classical setting. We illustrate that in the Appendix \ref{app:example_t_test}. 
\section{Shortcomings of Prior Work}\label{sec:shortcomings}

\begin{table*}
    \small
    \centering
    \caption{Comparison of previous approaches to \gls{irtlf}.}
    \begin{tabular}{lccccccccc}
        \toprule
        \textbf{Tool} & \makecell[c]{\textbf{Dependent}\\ \textbf{Data}}&\makecell[c]{\textbf{Non-}\\ \textbf{Parametric}}&\makecell[c]{\textbf{Bounded}\\ \textbf{type-1 error}} &
        \makecell[c]{\textbf{Arbitrary}\\ \textbf{Alternatives}}
        & \textbf{Bootstrap} & \makecell[c]{\textbf{No multiple}\\ \textbf{testing}} &  \makecell[c]{\textbf{Discrete} \\ \textbf{Data}}&  \makecell[c]{\textbf{Reservable} \\ \textbf{Hypotheses}}\\
        \midrule
        \gls{tvla}
        & \faAdjust  & \faCircle[regular] & \faCircle & \faCircle[regular] & \faCircle[regular] & \faCircle & \faCircle &\faCircle[regular] \\
        \gls{mona}
        & \faCircle[regular] &  \faCircle & \faCircle[regular] & \faCircle[regular] & \faCircle[regular] & \faCircle[regular] &  \faCircle[regular]&\faCircle[regular]\\
        \gls{dudect}
        &\faCircle[regular] &  \faCircle[regular] & \faAdjust & \faCircle[regular] & \faCircle[regular] & \faCircle[regular] &   \faCircle&\faCircle[regular]\\
        \gls{tlsfuzzer}
        & \faAdjust &  \faAdjust & \faCircle & \faCircle[regular]  &  \faAdjust & \faCircle[regular]  &   \faCircle &\faCircle[regular]\\ 
        \gls{rtlf}
        & \faCircle[regular] &  \faCircle & \faCircle & \faCircle[regular] & \faCircle  & \faCircle[regular] & \faCircle[regular]&\faCircle[regular] \\
        \gls{irtlf} 
        & \faCircle & \faCircle & \faCircle & \faCircle & \faCircle & \faCircle & \faCircle &\faCircle \\
        \midrule
        \multicolumn{9}{c}{\faCircle~Applies \hspace{5mm} \faAdjust~Partially applies \hspace{5mm} \faCircle[regular]~Does not apply}\\
        \bottomrule
    \end{tabular}
    \label{table:shortcomings}
\end{table*}

Previous approaches for the analysis of timing side channels suffer from various drawbacks, which we point out in this section. We summarize them in \autoref{table:shortcomings}.
 \paragraph{Parametric Assumption} As discussed before, parametric assumptions are often not valid in the context of timing side channels. The assumption of a normal distribution made by \gls{tvla}, \gls{dudect}, and also \gls{tlsfuzzer} (as it employs a paired two-sample t-test), limit their applicability for this use case. Note that in other fields, for example when comparing power traces for hardware leaks, parametric assumptions can be valid.
\paragraph{Bounded Type-1 Error}
As discussed in \autoref{sec:hypothesistest}, a hypothesis test is usually constructed in the spirit that for the bounded type-1 error, we want to minimize the type-2-error. However, neither Mona nor Dudect fully controls the type-1 error. Due to the construction of \gls{mona}, especially for small sample sizes and large variances, \gls{mona} faces an overwhelming type-1 error (see \cite{USENIX:DMEMBS24}). In contrast, \gls{dudect} uses a very conservative decision rule, which is the other extreme to \gls{mona} (see \cite{USENIX:DMEMBS24}).

\paragraph{Arbitrary Alternatives}
In practical applications, the ability to detect violations of $H_0$ (i.e. statistical power) is critical. Ideally, we require that $\mathbb{P}(\text{reject } H_0 \mid H_1 \text{ is true}) \overset{n \to \infty}{\to} 1$. 
However, this guarantee fails for some of the commonly used tools. For example, \gls{mona} relies on a fixed threshold and exhibits no statistical power when the differences between distributions fall below those thresholds. In the case of \gls{mona}, effects below a $1\%$ difference remain undetected, even for large sample sizes. Similarly, \gls{dudect} and \gls{tlsfuzzer} lack statistical power in very simple examples like two normal distributions with equal means but differing variances, highlighting the limitations of these tools. In analogy, \gls{rtlf} considers only differences in deciles, making it insensitive to changes that do not affect those specific quantiles. Extending their methodology to a larger collection of quantiles is possible but incurs severe multiple-testing costs.

\paragraph{Multiple Testing}
Mona, dudect, tlsfuzzer and RTLF use multiple hypothesis testing (see Section \ref{sec:background}). 
While \gls{dudect} and \gls{mona} do not explicitly account for this issue, \gls{rtlf} applies a simple Bonferroni correction \cite{bonferroni}, to control the overall level $\alpha$, while \gls{tlsfuzzer} attains a type-1 error of $4\alpha$. Even though the Bonferroni correction can be optimized in various ways (e.g., \cite{Sidak1967}, \cite{Holm1979}), there have been, to the best of our knowledge, no efforts to apply these methods in the side-channel literature. 
\paragraph{Dependent Data}
As already mentioned, dependence between observations is a natural phenomenon in real systems. To this day, no statistical tool has accounted for this data characteristic. Strictly speaking, the results obtained by all tools did not yield the desired guarantees. However, there is a grey zone in which minor deviations do not affect the outcome. This can, for example, be observed in the OpenSSL data collected in Dunsche et al.~\cite{USENIX:DMEMBS24}. While minor dependence between observations is present, the results are still sound. However, as we will later see (\autoref{sub:mbedtls}), the dependence between observations will often not be negligible.

\begin{figure*}[t]
    \centering
     \begin{subfigure}{0.195\textwidth}
        \centering
        \includegraphics[width=0.85\linewidth]{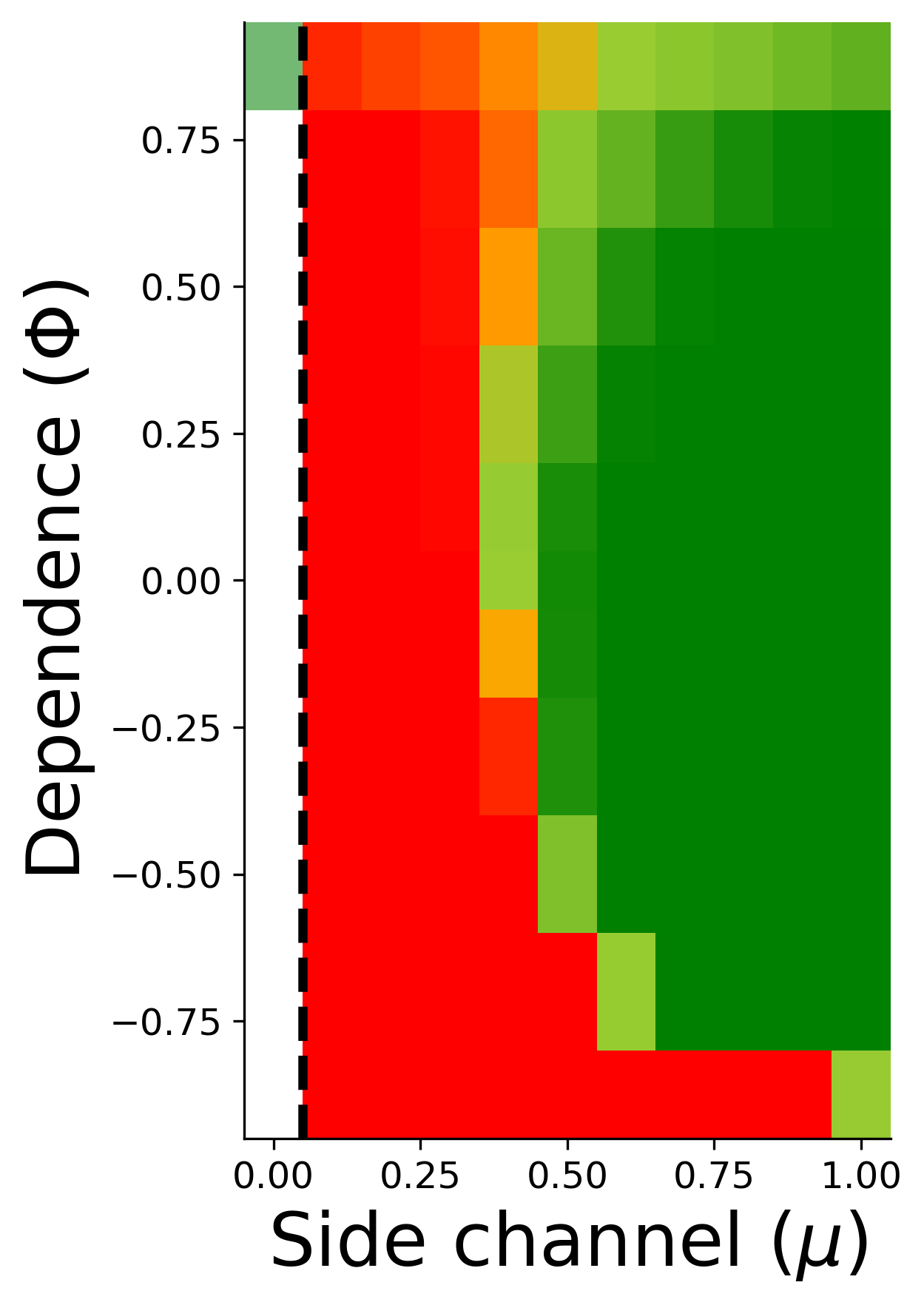}
        \caption{\centering\gls{dudect}}
        \label{fig:dudect}
    \end{subfigure}
    \begin{subfigure}{0.195\textwidth}
        \centering
        \includegraphics[width=.85\linewidth]{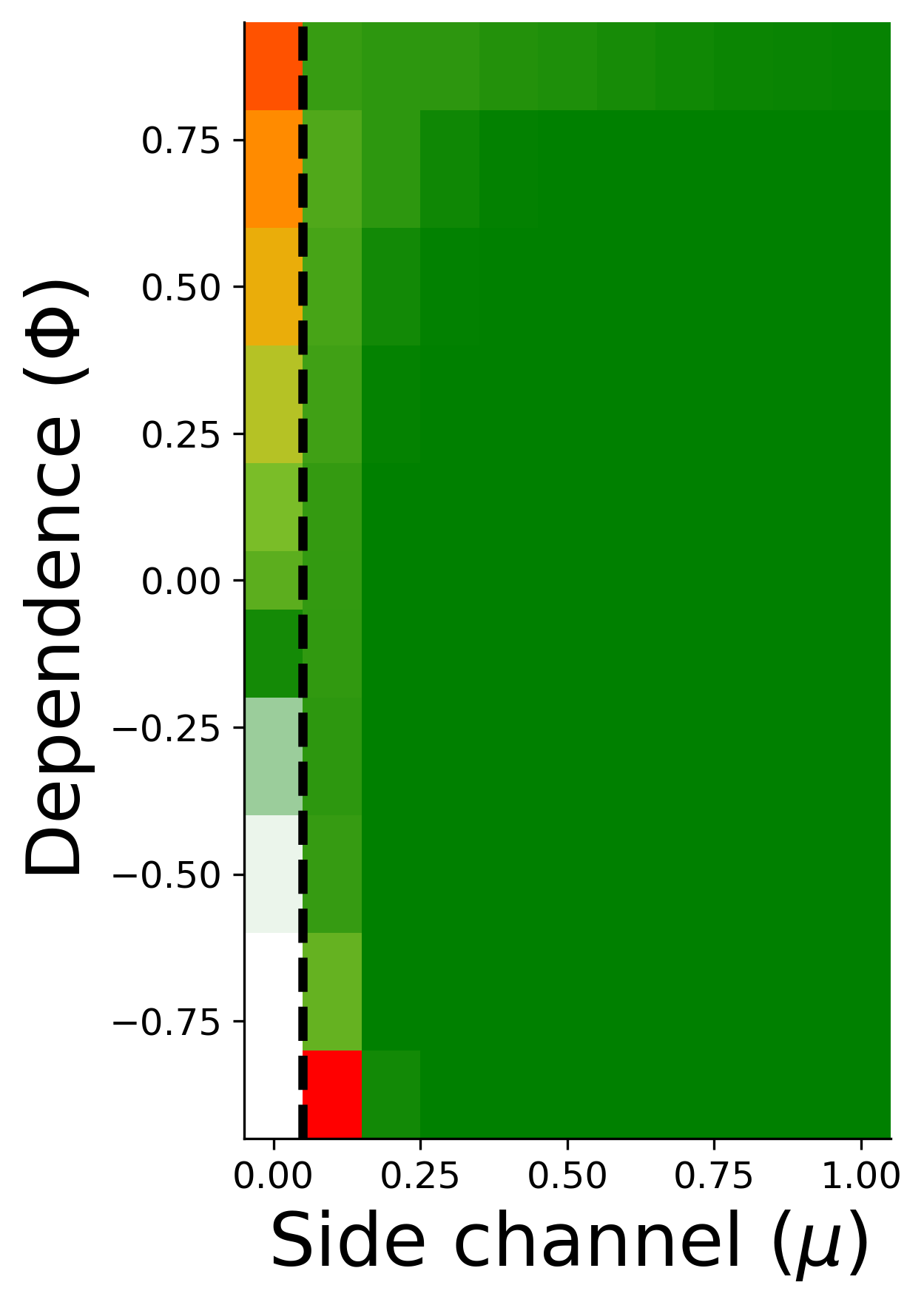}
        \caption{\centering \gls{tvla}}
        \label{fig:TVLA}
    \end{subfigure}
    \hfill
    \begin{subfigure}{0.195\textwidth}
        \centering
        \includegraphics[width=.85\linewidth]{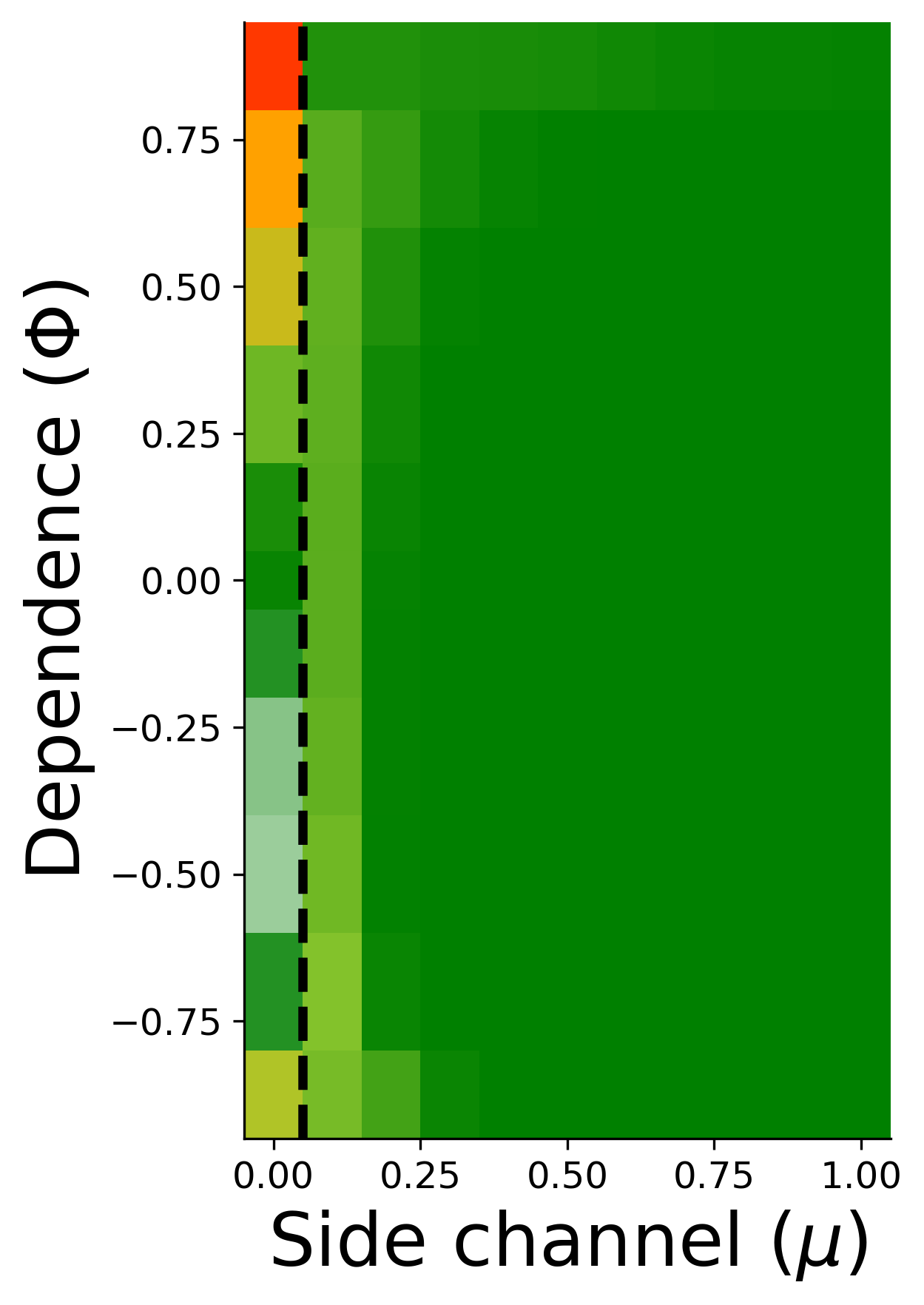}
        \caption{\centering \gls{rtlf}}
        \label{fig:RTLF}
    \end{subfigure}
    \hfill
    \begin{subfigure}{0.195\textwidth}
        \centering
        \includegraphics[width=0.85\linewidth]{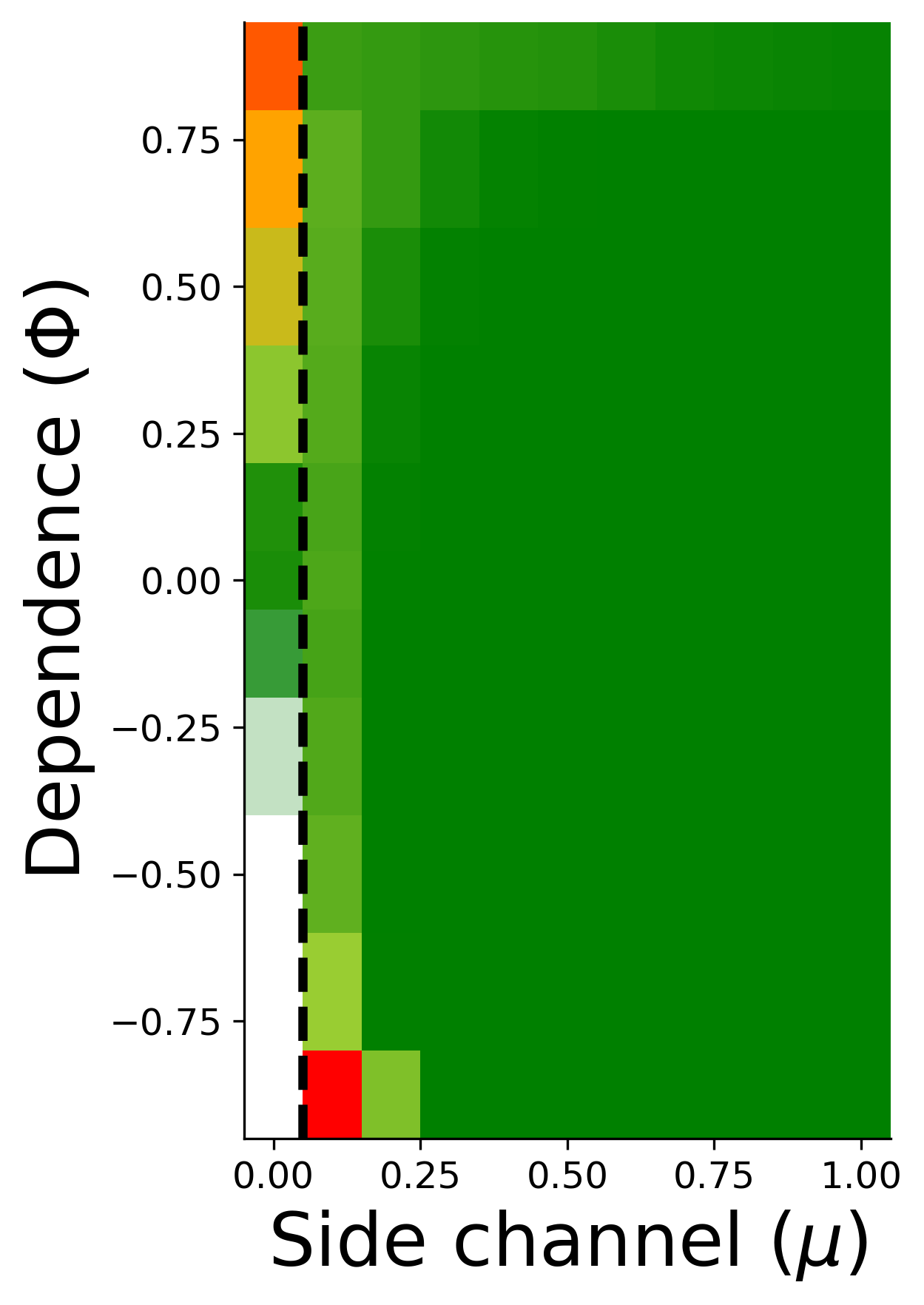}
        \caption{\centering \gls{tlsfuzzer}}
        \label{fig:tlsfuzzer}
    \end{subfigure}
    \hfill 
    \begin{subfigure}{0.195\textwidth}
        \centering
        \includegraphics[width=.5\linewidth]{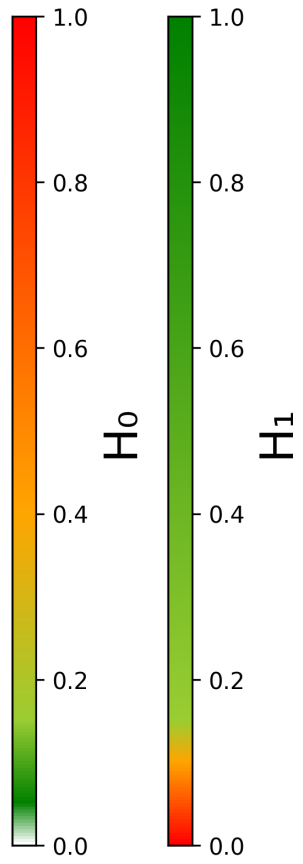}
    \end{subfigure}
    \caption{Empirical rejection rates of $1,000$ simulation runs for various tools with sample size $n=10000$. For \gls{tvla}, \gls{rtlf} and \gls{tlsfuzzer} we set $\alpha=0.1$, while for \gls{dudect} we kept the default threshold, which yields a very conservative $\alpha$.}
    \label{fig:heatmaps-other-tools}
\end{figure*}

We illustrate the impact of dependent data in a simple example, where we artificially generate dependent data for which we can accurately control the ground truth (see Appendix \ref{app:gen_data}). We then analyze the data with \gls{dudect}, Welch's t-test (\gls{tvla}), \gls{tlsfuzzer} (excluding the Friedman test as we consider a two-sample problem), and \gls{rtlf} (\autoref{fig:heatmaps-other-tools}). Our considered side channel uses a simple shift. Here, we want to emphasize that the designed experiments favor tools based on the t-test (\gls{tvla}, \gls{dudect} and \gls{tlsfuzzer}) since the data example recovers the comparison of two normally distributed samples with a shift for the independent case.
In this simple example, we see that all tools struggle in multiple ways. Starting with the null hypothesis ($H_0$), shown in the first column and separated by the black dotted line, we notice that all tools tend to over-reject for positively correlated measurements. Among them, only \gls{dudect} avoids extreme over-rejection, mainly because it employs a very conservative threshold. The statistical reason for this is that for positively correlated data, the estimated variance is smaller than the true variance, leading to too many false positives. 
In contrast to that, we can observe that for negatively correlated data, the exact opposite happens, namely under-rejection, due to an overestimation of the true variance. For the independent case, where $\Phi=0$, we can observe that all tests perform reasonably well. Moving to the alternative hypothesis, i.e. where a side channel is indeed present. Besides \gls{dudect}, all tools reliably detect side channels. Again, the conservative decision rule proposed by \gls{dudect} is the crucial part here. For positively correlated data under $H_1$, the detections are, in theory, valid. However, since the type-1 error ($\alpha$) was not properly controlled under $H_0$, drawing a usable conclusion is not possible. With strong positive dependence, distinguishing between the presence or absence of a side channel becomes particularly difficult, as a positive outcome (side channel present) is as likely under $H_0$ as in $H_1$. For negative correlations, there is an actual turning point. If the signal $\sqrt{n}\mu$ surpasses the threshold $c$ used by the respective tool, an under-rejection becomes an over-rejection. This is clearly visible in \gls{tvla} and \gls{tlsfuzzer} from the second to the third column and less obvious in \gls{dudect} and \gls{rtlf}.


\section{Real-World Examples}
\paragraph{Dependent Data}
Dependent observations naturally occur in measurement data, as each measurement on real hardware has the possibility to influence subsequent measurements as the systems we are measuring on are \emph{stateful}. Each measurement influences the values in CPU registers, the RAM, the OS, and various caching mechanisms. On a micro-architectural level, the measurement can also influence the power consumption~\cite{USENIX:WPHSFK22}, which influences the CPU frequency. These changes in the system state can create a dependency in the measurements that is measurable. This dependency can be very small if functions are measured in isolation, where side effects can be well contained, or can be very large if side channels are measured within bigger applications that do more than just execute a mathematical function in a tight loop. To show that data dependency can naturally occur, we conducted an experiment built around the page table mechanism in the operating system and CPU. When a process requests memory from the operating system, a piece of physical memory is mapped to a virtual address space. This mapping takes heavy advantage of a cache called TLB (Translation Lookaside Buffer). If a piece of memory (page) was recently loaded, it will be in the cache, and access to it will be faster, while not-so-recent pages will be slower since they are not in the cache anymore and need to be loaded again. Even if the information that is retrieved is public (and the timing difference, therefore, not an issue), the TLB can introduce data dependency, which depends on the memory access pattern of the inputs in the system under test. To demonstrate this effect, we implemented an experiment where we allocate a large piece of memory (32MB) and then either access the first 512KB of memory or a random chunk of 512 KB. When frequently accessing the first 512 KB, we expect the respective pages to be in the TLB, and the measurements should, therefore, not have a strong dependency, as the pages do not have a chance to unload between measurements. However, when accessing random chunks of memory, the access time will depend on whether the chunk, or parts of it, were accessed by a previous (random) measurement. It will, therefore, create a stronger dependency, as the caches are eventually purging pages from the cache. In an experiment with 100,000 measurements repeated 10,000 times, we found that the dependency analysis revealed an average dependency estimation of 29.35 when accessing the same chunk, while the average dependency estimation was 132.15 when accessing random chunks. Here, higher values correspond to longer dependence between measurements. For more details, we refer to Section \ref{sec:hypothesis_test}. 
In \autoref{sec:real_world}, we present an example from Dunsche et al. \cite{USENIX:DMEMBS24} where during the Bleichenbacher vulnerability test in \gls{mbedtls}, data dependency appeared in real-world measurements, which was likely responsible for a false positive result in \gls{rtlf}.
\paragraph{Negligible Side Channels}
Developers always consider (sometimes implicitly) specific attacker scenarios as out of scope. For example, OpenSSL, the most well-known cryptographic library, does not consider attacks that require co-located attackers to be relevant.\footnote{\url{https://openssl-library.org/policies/general/security-policy/}} In the context of timing measurements, this may apply to known but small side channels that likely can not be exploited in a remote or even LAN attacker scenario with a realistic amount of measurements. Through our methodology, we provide a tool to developers that allows them to define their notion of a \textit{negligible} side channel. This way, our statistical test can be included in a continuous integration pipeline that detects any new side channels exceeding the defined threshold while maintaining a low false positive rate for the expected leak. Additionally, setting thresholds enables testers to produce more meaningful reports as they can focus on reporting exploitable side channels.
We demonstrate the effect of our threshold in \autoref{sec:real_world} based on a known side channel in an older version of the reference code of the post quantum KEM scheme Kyber~\cite{bernstein2025kyberslash}.
\paragraph{Determining a Reasonable Sample Size}
The ability to estimate a required sample size for a reliable detection of timing leaks can be a beneficial feature for developers. However, the required sample size is heavily influenced by measurement variance. Our approach allows for a sample size estimation by analyzing a small initial sample to assess variance, then incorporating the detection threshold ($\Delta$), critical leak size ($\mu$), and desired detection rate ($p$). \autoref{sec:real_world} demonstrates this estimation on real-world data.
\section{\gls{irtlf}: Relevant Hypothesis Tests}
\label{sec:hypothesis_test}

In this section, we propose a statistical tool called SILENT (Statistical Identification and Leakage Estimation with Negligibility Thresholds) to distinguish between two distributions $P_X$ and $P_Y$ based on collected timing measurements of $X$ and $Y$ using the hypotheses \eqref{det1}. Our test does not rely on parametric assumptions, will bound type-1 errors, and support arbitrary alternatives with minor additional costs. The test decision is based on a bootstrap procedure. We first define some basic assumptions required for statistical theory which confirms the feasibility of the proposed method:
\begin{As}\label{Assumptions}
  Let $Z_i = (X_i, Y_i)^\top$, $i = 1, \ldots, n$, $n \in \mathbb{N}$, be a sequence of strictly stationary $m$-dependent random vectors for some $m \in \mathbb{N}$.
\end{As}
 Here, we emphasize that we do not assume independence between $X_i$ and $Y_i$, even though this can be the case in practice. Additionally, we allow for dependence between the vectors $Z_i$ in a specific manner, which is called  {\it $m$-dependence}  and defined as follows: for a fixed $m \in \mathbb{N}$ and any $t \in \mathbb{N}$, the sequence $(Z_i)_{i \leq t}$ is independent of $(Z_i)_{i \geq t + m + 1}$. Note that this recovers the case of independence when $m = 0$. Obviously, $m$ is unknown in advance and therefore we estimate it using the generic estimator proposed by Politis et al.~\cite{Politis31122004}. An implementation can be found in ~\cite{bstar}.
Beyond this, we impose no additional distributional assumptions, particularly no parametric assumptions. Recall that the goal of the statistical hypothesis test is to decide whether the distributions $P_X$ and $P_Y$ are similar or not. To achieve this, we construct a hypotheses test based on a slack parameter $\Delta > 0$, which will be chosen in advance to determine whether two distributions are "$\Delta$-close" or not.  

Subsequently, we compare $P_X$ and $P_Y$ using a set of quantiles of $X$ and $Y$, respectively. For that, we first define a set of indices $K$, indicating which quantiles will be considered. We consider the following hypotheses-pair
\begin{equation}\label{eq:null_hyp}
    H_0: \max_{k\in K}|q_k^X-q_k^Y|\leq \Delta
\end{equation}
versus
\begin{equation}\label{eq:alt_hyp}
    H_1:\max_{k\in K}|q_k^X-q_k^Y|>\Delta~.
\end{equation}
Note that whenever the alternative hypothesis $H_1$ holds, the two distributions are indeed different.
\paragraph{Maximum test}
In contrast to previous approaches, we will not use a multiple testing approach, but rather test all quantiles simultaneously. For that purpose, we first define the test statistic
\begin{equation}\label{eq_test_statistic_max}
  \hat Q_{max}:=   \hat Q_{max}(x,y):=  \max_{k\in K_{sub}^{max}}\frac{|\hat q_k^x-\hat q_k^y|-\Delta}{\hat\sigma_k}~,
\end{equation}
where $\hat\sigma_k^2$ is the empirical variance of the bootstrap sample\linebreak
$\hat Q_1^{k,*}, \hdots , \hat Q_B^{k,*}$ calculated by Algorithm \ref{algorithm:bootstrap_max}), i.e.
\begin{equation}\label{eq:variance}
\hat\sigma_k^2:= \frac{1}{B-1}\sum_{i=1}^B (\hat Q_i^{k,*} - \bar Q_i^{k,*})^2~, 
\end{equation}
and  $\bar Q_i^{k,*}:=\frac{1}{n}\sum_{i=1}^n \hat Q_i^{k,*}$ denotes its empirical mean.
Regarding the definition of $K_{sub}^{max}$, we filter $K$ twice. In the first step, we define the subset
 \begin{equation}\label{eq:subset_sub}
K_{sub}:= \Big\{k\in K \Big|  
 \hat\sigma_k^2< 5 \frac{1}{|K|}\sum_{k\in K} \hat \sigma_k^2\Big\}  
 \end{equation}
 of quantiles, where the variance is not too large compared to the rest. This also excludes the scenario of unfavorable signal-to-noise ratios, and we include only those quantiles $q_k$, which can be estimated with a sufficiently large number of observations. In the second step, we also define  the subset
   \begin{equation}\label{eq:subset_sub_max}K_{sub}^{max}:= \Big\{k\in K_{sub} \Big| \frac{|\hat q_k^X-\hat q_k^Y|}{\hat\sigma_k}+30 \sqrt{\frac{\log(n)^{3/2}}{n}} \geq \frac{\Delta}{\hat\sigma_k}\Big\}
   \end{equation}
    as the set of levels with relevant differences between the quantiles. This filter is primarily used for boosting the statistical power of the resulting test, as it includes only those differences between the quantiles that are likely to exceed $\Delta$. Note that this can also yield less quantiles in the bootstrap procedure and therefore a less conservative decision rule.
    With that in hand, we can define the test decision by the following: we reject the null hypothesis \eqref{eq:null_hyp} in favor of \eqref{eq:alt_hyp}, if and only if 
    \begin{equation*}
        \hat Q_{max}(x,y)-{c^*_{1-\alpha}}>0~,
    \end{equation*}
    where ${c^*_{1-\alpha}}$ is the threshold determined by the resampling procedure in Algorithm \ref{algorithm:test} and Algorithm \ref{algorithm:bootstrap_max}. Here, we point out that the procedure defined in Algorithm \ref{algorithm:bootstrap_max} refers to the continuous case. For the discrete case, the bootstrap procedure in Algorithm \ref{algorithm:bootstrap_max} is not consistent. Therefore, we have formulated an alternative in Algorithm \ref{algorithm:bootstrap_max_discrete} based on Jentsch and Leucht~\cite{bootstrap_discrete}.

\paragraph{Quantile Estimator}
We propose the following estimators: For continuous data, we propose the well-known estimator based on the rank statistic: 
\begin{equation}\label{quant_statistic}
    \hat q_i^x:= \begin{cases} x_{(\lceil n i\rceil)}, &\text{if} \quad ni\notin \mathbb{Z}\\ \frac{1}{2} (x_{(ni)}+ x_{(ni+1)}), &\text{if} \quad ni\in \mathbb{Z}
    \end{cases}~.
\end{equation}
For discrete data, this estimator is not usable, as the classical asymptotic analysis (see e.g. \cite{vaart1998}) is not applicable.  Therefore, we consider a different estimator for discrete distributions, so-called mid-distribution quantiles (see Appendix \ref{app:add_def_alg}).

\subsection{Algorithms}
The algorithmic foundation comprises three key components: calculating the test statistic defined in \eqref{eq_test_statistic_max}, determining the threshold, and making the final test decision. Algorithm \ref{algorithm:test} integrates all three steps. Since threshold computation is a crucial contribution of this work, we present it separately in Algorithm \ref{algorithm:bootstrap_max} (or Algorithm \ref{algorithm:bootstrap_max_discrete} for discrete data). We implemented our algorithm in R and make it publicly available on GitHub\footnote{\url{https://github.com/tls-attacker/SILENT}}. For the implementation we have used the following packages: \cite{Qtools,lrv,bstar}.

\begin{algorithm}
    \small
          \begin{flushleft}\textbf{function} \textsc{\gls{irtlf}}($x$, $y$, $\alpha$, $\Delta$, $K$, $B$)\end{flushleft}
    \begin{algorithmic}[1]
    \REQUIRE data sets $x$, $y$, false positive rate $\alpha$, threshold $\Delta$, set of quantiles $K$, bootstrap parameter $B$
    \ENSURE "Violation", "No Violation"
    \STATE Check if data is discrete or continuous.
    \STATE Compute quantiles $\hat q^x$ and $\hat q^y$.
    \STATE Estimate dependence $m$.
    \STATE Compute bootstrap statistics $\hat Q^* := Bootstrap(x,y,K, B, m)$.
    \STATE Compute empirical variance $\hat \sigma_k^2$ (as in equation \eqref{eq:variance}).
    \STATE Select subset $K_{sub}^{max}$ defined in equation \eqref{eq:subset_sub_max}.
    \STATE Compute test statistic $\hat Q_{max}(x,y)$ as in equation \eqref{eq_test_statistic_max}
    \STATE Compute final bootstrap statistic $\hat Q_{max}^* := \max_{k \in K_{sub}^{max}} \hat Q^{k,*}/\hat\sigma_k$
    \STATE Define threshold $c_{1-\alpha}^* := (\hat Q_{max}^*)_{\lfloor(1-\alpha)B\rfloor}$
    \IF{$\hat Q_{max}(x,y) > c_{1-\alpha}^*$}
        \STATE \textbf{return} "Violation".
    \ELSE
        \STATE \textbf{return} "No Violation".
    \ENDIF

    \end{algorithmic}
    \caption{\gls{irtlf} Test to distinguish two distributions.}
    \label{algorithm:test}
\end{algorithm}

\begin{algorithm}
    \small
    \begin{flushleft}\textbf{function} \textsc{Bootstrap}($x$, $y$, $K$, $B$, $m$)\end{flushleft}
    \begin{algorithmic}[1]
    \REQUIRE data sets $x$, $y$, set of quantiles $K$, bootstrap parameter $B$, dependence $m$
    \ENSURE $\hat Q^*$
    
    \FOR{$i = 1$ to $B$}
        \STATE Sample $I \subset \{1, \hdots, n - m + 1\}$ with $|I| = \lceil n/m \rceil$.
        \STATE Set $x^* = x[I],\ y^* = y[I]$, where each $i \in I$ includes all $x_i, \hdots, x_{i+m-1}$.
        \STATE Compute bootstrap test-statistic:
        \[
        \hat Q^{i,*} = \left| \hat q^{x^*} - \hat q^{y^*} \right| - \left| \hat q^x - \hat q^y \right|~.
        \]
    \ENDFOR

    \STATE \textbf{return} $\hat Q^* := (\hat Q_1^{*}, \hdots, \hat Q_B^{*})$.
    \end{algorithmic}
    \caption{Bootstrap maximum for dependent data}
    \label{algorithm:bootstrap_max}
\end{algorithm}

\begin{algorithm}
    \small
    \begin{flushleft}\textbf{function} \textsc{Measurements}($x$, $y$, $\mu$, $\Delta$, $K$, $Shift$,$\alpha$)\end{flushleft}
    \begin{algorithmic}[1]
    \REQUIRE data sets $x$, $y$, expected side channel $\mu$, threshold $\Delta$, set of quantiles $K$, boolean $Shift$, type-1 error threshold $\alpha$
    \ENSURE $n$
    \STATE Compute quantiles $\hat q^x$ and $\hat q^y$.
    \STATE Estimate dependence $m$.
    \STATE Compute bootstrap statistics $\hat Q^* := Bootstrap(x,y,K,B,m)$.
     \STATE Compute empirical variance $\hat \sigma_k^2$ (as in equation \eqref{eq:variance}).
    \STATE Select subset $K_{sub}$ as in \eqref{eq:subset_sub}.
   \IF{$Shift = 1$}
        \STATE $\hat\sigma:= \sqrt{n}\min_{k\in K_{sub}}\hat \sigma_k$.
    \ELSE
        \STATE $\hat\sigma:= \sqrt{n}median(\hat \sigma_k)$.
    \ENDIF
    \STATE Compute $n_{sub}=\Big(\frac{\frac{\Phi^{-1}(1-p)}{\hat\sigma}-\Phi^{-1}(1-\alpha)\hat\sigma}{\mu-\Delta}\Big)^2$, where $\Phi^{-1}(\cdot)$ is the quantile function of a standard normal distribution.
    \STATE \textbf{return} Estimated sample size $n=\lceil max(100, n_{sub})\rceil$.
    \end{algorithmic}
    \caption{Statistical Power Analysis}
    \label{algorithm:power_analysis}
\end{algorithm}

\subsection{Analysis under the Null Hypothesis}\label{subsec:analysis_h_0}
Subsequently, we consider two different asymptotic results. Theorem \ref{thm:asymptotic} derives the limit on the boundary of $H_0$, namely $\max_{k\in K}|q_k^X-q_k^Y|=\Delta$. On the other hand, Theorem \ref{thm:bootstraplevel} derives the asymptotic $\alpha$-level for the bootstrap quantile.
\begin{thm}[Asymptotic behavior]\label{thm:asymptotic}
Assume that Assumption \ref{Assumptions} holds and $X$ and $Y$ are continuous random variables with densities $f_X,$ and $  f_Y$, which are strictly positive and continuous in a neighborhood of all points $k \in K$. 
Furthermore let $\mathbb{G}=(\mathbb{G}_1,\mathbb{G}_2)$ denote a bivariate Brownian bridge with covariance structure 
\begin{align}
\label{det10}
    \begin{pmatrix}
        \sigma_{X,l}^2 & \sigma_{X,Y,l}\\
        \sigma_{X,Y,l} & \sigma_{Y,l}^2
    \end{pmatrix}
    = \sum_{h =-m}^m  
    \begin{pmatrix}
        {\rm Cov} (X_0,X_h) &  {\rm Cov} (X_0,Y_h)\\
         {\rm Cov} (Y_0,X_h) &  {\rm Cov} (Y_0,Y_h)
    \end{pmatrix}.
\end{align}
If the null hypotheses \eqref{eq:null_hyp} is satisfied, the weak convergence   \begin{equation*}
        \hat Q_{max} \overset{d}{\to}\max_{k \in K, |q_k^X-q_k^y|=\Delta}\text{sign}(q_k^X-q_K^Y)\left(\frac{\mathbb{G}_1(k)}{f_X(q_k^X)}- \frac{\mathbb{G}_2(k)}{f_Y(q_k^Y)}\right)
    \end{equation*}
    holds,  where the right hand side is defined as   $-\infty$ if $\big \{ k \in K, |q_k^X-q_k^y|=\Delta \big \}= \emptyset $ . 
\end{thm}
In Theorem \ref{thm:asymptotic}, we only consider the continuous case. The discrete case is slightly more involved but yields no further insights. Asymptotic results for mid-distribution quantiles can be obtained in \cite{ma2011asymptotic}.
In the following theorem, we formulate the guarantee for our bootstrap procedure under the null hypothesis:
\begin{thm}[Bootstrap Guarantee]\label{thm:bootstraplevel}
Let the assumptions of Theorem \ref{thm:asymptotic} be satisfied and let $c^*_{1-\alpha}$ denote  the quantile obtained by  Algorithm \ref{algorithm:test}. Suppose that the null hypothesis \eqref{eq:null_hyp} holds,  then
\begin{align*}
    \limsup_{n \to \infty} \mathbb{P}( \hat Q_{max}
    > c^*_{1-\alpha})\leq \alpha
\end{align*}
with equality in the case that $|q_k^X-q_k^Y|=\Delta$ for all $k \in K$.
\end{thm}

\subsection{Analysis under the Alternative Hypothesis}\label{subsec:analysis_h_1}
To complement the results in Section \ref{subsec:analysis_h_0}, we also consider the alternative hypothesis $H_1$ in \eqref{eq:alt_hyp}. We first derive the consistency, which means that the probability of rejection converges to $1$ if the null hypothesis is not satisfied. Secondly, we also derive a "worst case" statistical power analysis. 
\begin{thm}[Bootstrap Guarantee]\label{thm:bootstrappower}
Let the assumptions of Theorem \ref{thm:asymptotic} be satisfied and 
 let $c^*_{1-\alpha}$ denote  the quantile obtained by  Algorithm \autoref{algorithm:test}.
Suppose that the alternative  \eqref{eq:alt_hyp} holds, 
then
\begin{align*}
    \lim_{n \to \infty} \mathbb{P}(\hat Q_{max}
    > c^*_{1-\alpha})=1
\end{align*}
In particular $\hat Q_{max}
$ diverges to infinity with rate $\sqrt{n}(\max_{k \in K}|q_k^X-q_k^Y|-\Delta)$. Suppose that there is only one quantile $q_{k_0}$ such that $|q_k^X-q_k^Y|$ is maximized; in this case the statistical power of the test is approximately given by 
\begin{align}\label{eq:power_single_alt}
    \mathbb{P}(Z>q_{1-\alpha}-\sqrt{n}(|q_{k_0}^X-q_{k_0}^Y|-\Delta))~,
\end{align}
where $Z$ is a centered, normally distributed random variable with variance
$$
\frac{\sigma_{F_X,l}^2}{f_X^2(q_{k_0}^X)}-\frac{2\sigma_{F_X,F_Y,l}^2}{f_X(q_{k_0}^X)f_Y(q_{k_0}^Y)}+\frac{\sigma_{F_Y,l}^2}{f_Y^2(q_{k_0}^Y)}~,
$$
%
$q_{1-\alpha}$ is the $(1-\alpha)$ quantile of its distribution and $\sigma_{F_X,l}^2, \sigma^2_{F_X,F_Y,l}$ and $\sigma_{F_Y,l}^2$ are the long-run (co)variances defined in \eqref{det10}.
\end{thm}

\section{Illustration of Contributions}
\label{sec:illustration_of_contribution}
In this section, we illustrate the contributions made in Section \ref{sec:hypothesis_test} on artificially created data. To do this, we first show that we have constructed a well-performing test illustrated on the same data sets as in Section \ref{sec:shortcomings} (see Appendix \ref{app:gen_data} for details) and later expand on the individual contribution of a statistical power analysis.

\begin{figure*}[t]
    \centering
    \begin{subfigure}{0.195\textwidth}
        \centering
        \includegraphics[width=.85\linewidth]{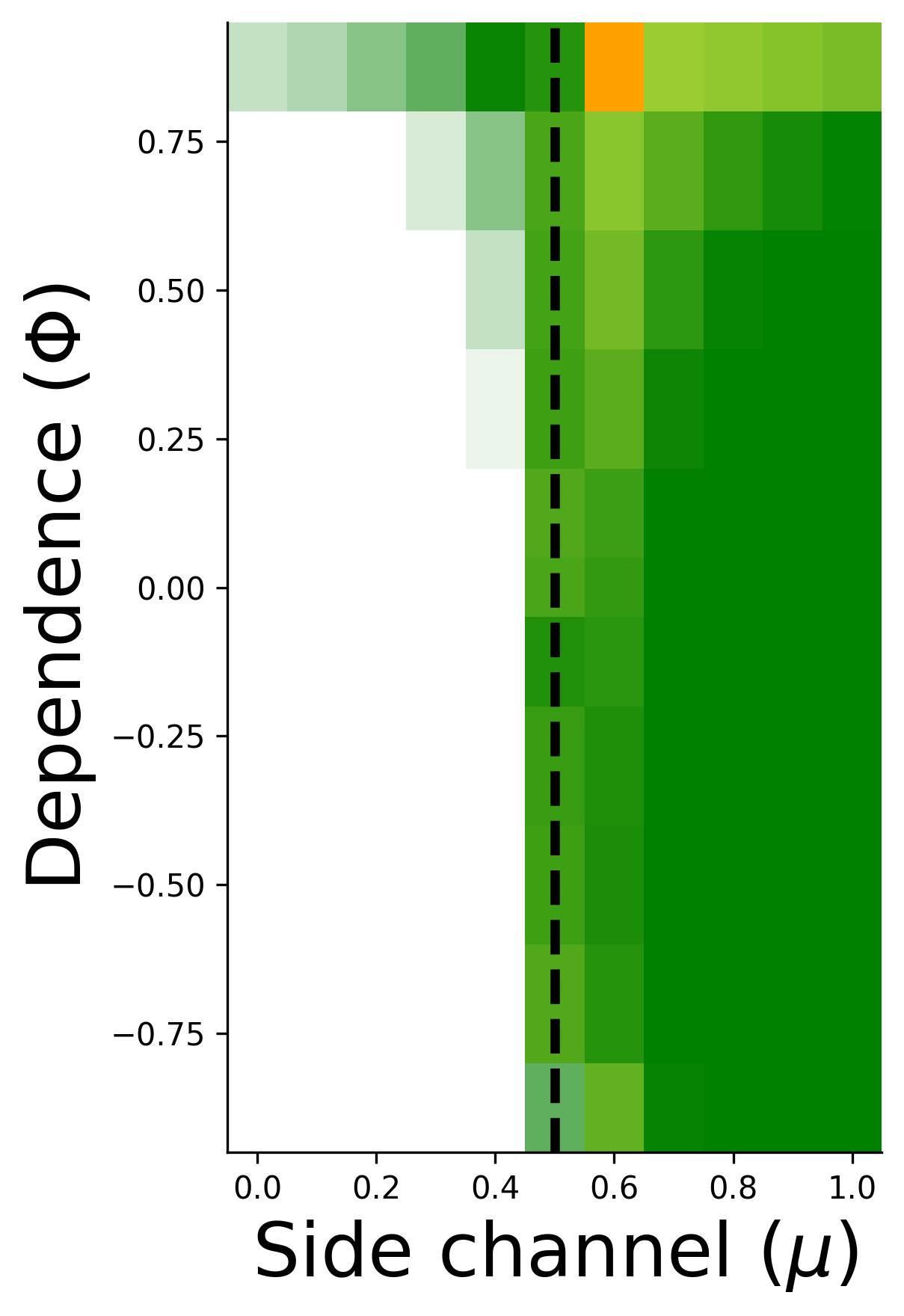}
        \caption{$n=10000$, $\Delta=0.5$}
        \label{fig:NEW_RTLF_10000}
    \end{subfigure}
    \hfill
    \begin{subfigure}{0.195\textwidth}
        \centering
        \includegraphics[width=.85\linewidth]{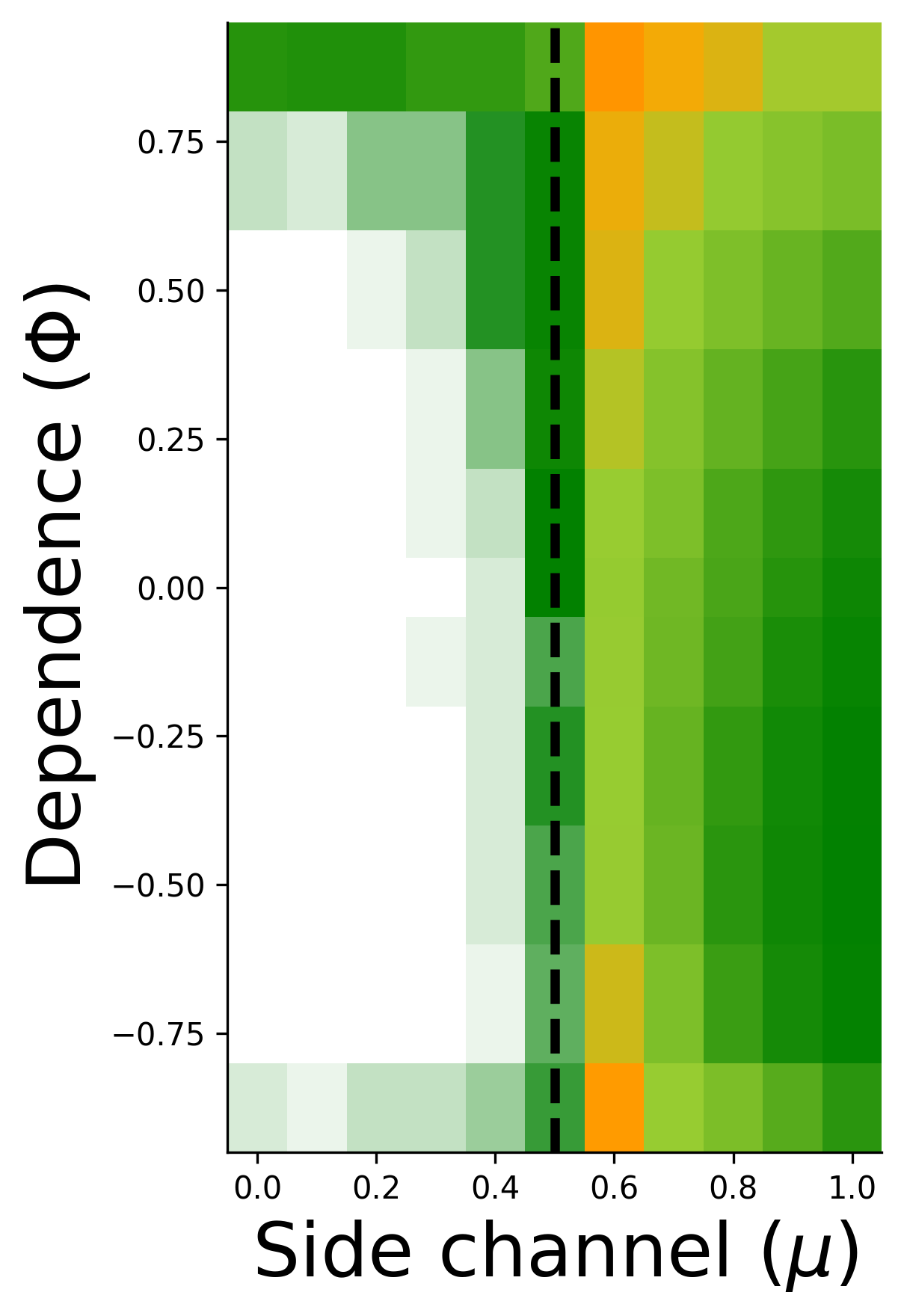}
        \caption{ $n=1000$, $\Delta=0.5$}
        \label{fig:NEW_RTLF_1000}
    \end{subfigure}
    \hfill
    \begin{subfigure}{0.195\textwidth}
        \centering
        \includegraphics[width=0.85\linewidth]{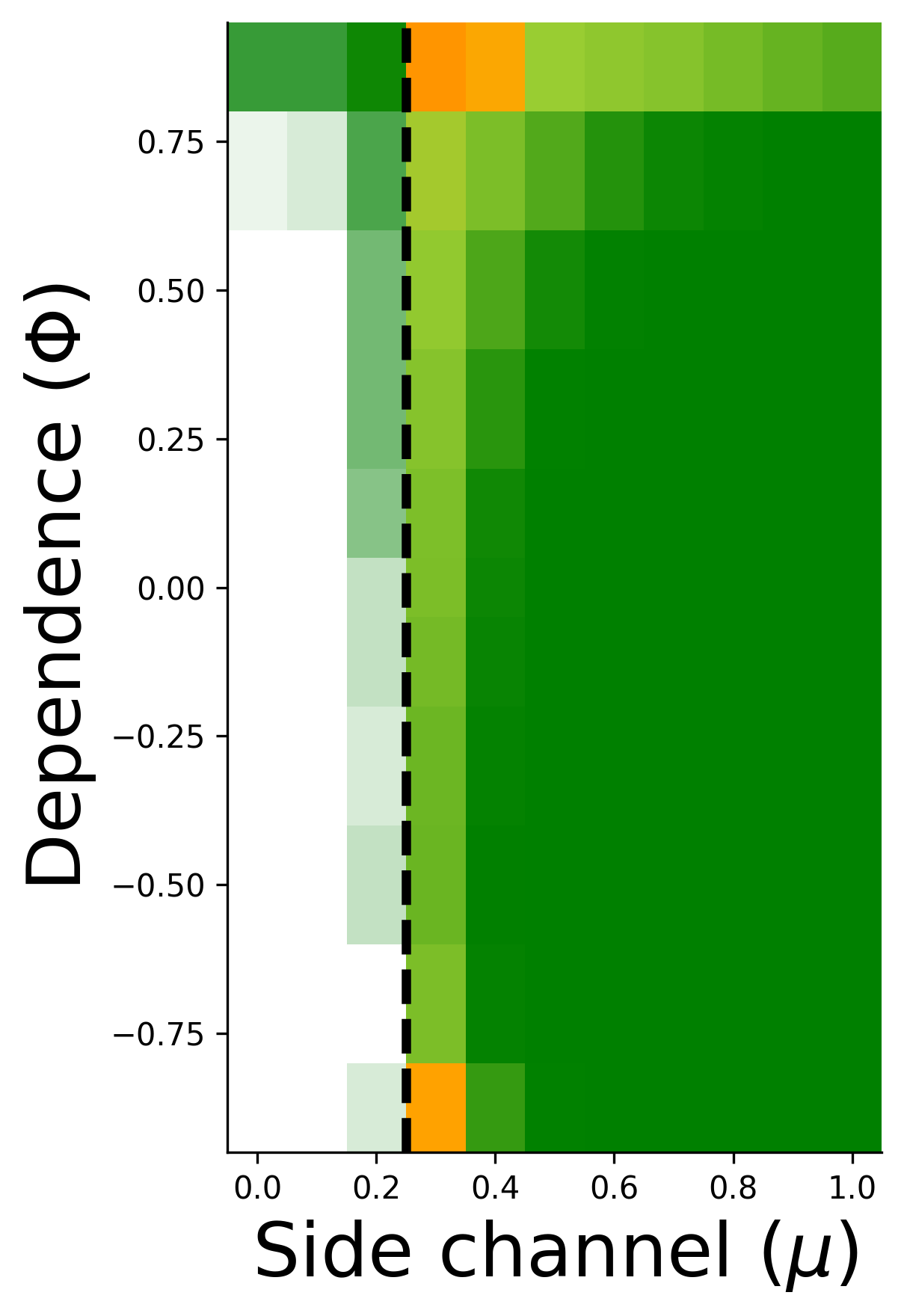}
        \caption{ $n=10000$, $\Delta=0.25$}
        \label{fig:NEW_RTLF_10000_delta}
    \end{subfigure}
    \hfill 
    \begin{subfigure}{0.195\textwidth}
        \centering
        \includegraphics[width=.5\linewidth]{figures/legend.png}
    \end{subfigure}
    \caption{Rejection rate heatmaps for \gls{irtlf} under varying sample size and $\Delta$.}
    \label{fig:heatmaps-nptd}
\end{figure*}

\paragraph{Varying Parameters}
We consider three different scenarios to illustrate the impact of $\Delta$ and sample size $n$: $n=1000, 10000$ with $\Delta=0.5$ and $n=10000$ with $\Delta=0.25$.
The rest of the parameters are set to default values. The results are summarized in Figure \ref{fig:heatmaps-nptd}.

In Figure \ref{fig:NEW_RTLF_10000} we revisit the example discussed in Section \ref{sec:shortcomings}. In contrast to all previously considered tools, we observe that \gls{irtlf} consistently maintains the type-1 error rate $\alpha$ across all forms of dependence $\Phi$. Notably, for $\mu < \Delta$, we clearly see the convergence $\alpha \to 0$ as $n \to \infty$. This is a highly relevant result: when the implementation is truly constant time (i.e., $\mu = 0$), the test yields practically no false positives and therefore significantly reduces the overhead created by investigation of false alarms. We refer the interested reader to Appendix \ref{app:example_t_test} where we explain how this improves over the naive two step procedure, where one first applies a classical test (e.g. one of the tools in \autoref{table:shortcomings}) and discards all positive test results where the detected difference is smaller than $\Delta$. Furthermore, Figure \ref{fig:NEW_RTLF_10000} also illustrates that if there indeed is  a side channel greater than $\Delta$ present, the  statistical power of the testing procedure quickly increases with the amount of measurements. This illustrates our ability to account for multiple testing penalties without sacrificing too much statistical power which is a classical problem of the Bonferroni correction that we evade due to our usage of $K_{sub}^{max}$ in Algorithm \ref{algorithm:test}. Turning to Figure \ref{fig:NEW_RTLF_1000}, we observe similar behavior for smaller sample sizes. While the test retains type-1 error control, its statistical power is reduced. The effect is strongest under strong positive correlation between measurements. This is expected, as such correlations increases the true standard deviation and thereby reduce the signal-to-noise ratio. This makes finding a true positive more difficult - or in simpler terms, with more noise for the same signal, detection becomes less reliable. 

Finally, Figure \ref{fig:NEW_RTLF_10000_delta} illustrates the flexibility in choosing different $\Delta$ values to obtain the theoretical guarantees established in Section \ref{sec:hypothesis_test}. Here, we set $\mu= 0.2$ and $\mu=0.3$, such that $\Delta=0.25$ lies between the two. This example highlights how suitable choices of $\Delta$ can effectively separate $H_0$ and $H_1$, leading to both type-1 and type-2 errors vanishing for sufficiently large $n$. In Figure \ref{fig:NEW_RTLF_10000_delta}, due to strong dependence and an insufficient sample size $n$, this ideal separation is not yet fully achieved. 

\paragraph{Statistical Power Analysis and Implications:} 

In practical settings, we often do not have prior knowledge of the precise noise characteristics present in measurements. Therefore, we expect users to collect at least $100$ measurements. Based on these, the expected size $\mu$, expected bias $\Delta$, desired detection rate $p$, and prescribed type-1 error $\alpha$, we will then conclude what the actual sample size $n$ should be to detect that difference $\mu$ reliably. An additional and equally important factor is the nature of the side channel. Specifically, is the leakage visible as a clear shift in the distribution, or is it only apparent in the tails—for instance, only in very fast execution times (e.g., the lower $10\%$ quantile)? This distinction has a direct impact on the sample size $n$. If the side channel causes a shift across the distribution, detection typically requires fewer samples. In contrast, when the effect is subtle and localized in the tails, more samples are needed to confidently identify the deviation.
To account for this, we implemented a simple but effective rule:
\begin{itemize}
    \item If the user believes the side channel is given by a shift, we estimate the variance using the smallest quantile variance observable by the initial measurements.
    \item If the user is unsure or suspects the effect may be local (say $10\%$ quantile only), we conservatively estimate using the median of the quantile variances observable by the initial measurements.
\end{itemize}
This will then also be reflected in the needed sample size $n$, as smaller variance yields smaller sample size.
\paragraph{Limitations} By measuring only small subsample to estimate the number of required measurements, we assume that the system under test is not meaningfully changing due to effects outside of our measurements. Strict stationarity is usually not given, as, for example, when doing network measurements, of course also, other processes on the system are changing the internal state of the system under test. However, for practical purposes there is some leniency in the requirement and subtle differences may even out over the course of the measurements. While the general order of magnitude of the statistical power analysis will be oftentimes still correct, better results are achieved when slightly more measurements are done when performing measurements on real systems.

\section{Real World Evaluation}
\label{sec:real_world}
In the following subsections, we analyze real-world measurements showcasing that our test can handle dependent data, the effect of $\Delta$ and how various parameters affect our sample size estimation, and the impact of different measurement setups.

\subsection{Dependent Data}
\label{sub:mbedtls}
First, we analyze a data set collected by Dunsche et al.~\cite{USENIX:DMEMBS24} for the mbedTLS library implementing the Transport Layer Security (TLS) protocol. Specifically, we analyzed their Bleichenbacher measurements in for the most recent version of mbedTLS they considered. We specifically chose this data set as their tool, RTLF, indicated an unexpected timing leak based on their measurements and we identified strong dependence between the measurements. As discussed in \autoref{sec:shortcomings}, RTLF may fail to maintain its configured false positive rate for such measurements. When analyzing the data set with our own test, we found no indication of a timing difference with the parameters $\alpha=0.1, \Delta=5$ and $B=1000$. As Dunsche et al.~\cite{USENIX:DMEMBS24} could not find a source for the perceived leak in the code of mbedTLS and already ruled this result to be a false positive, we believe that the dependency in the measurements likely caused the incorrect assessment made by RTLF. Another indicator is the statistical power analysis. For $\mu=100$ (as Dunsche et al. identified), $\Delta=5, p=0.9, \alpha=0.1$, we get an estimated sample size of $n=13,388,944$. This underpins that the detection in Dunsche et al. was likely a false positive.

\subsection{Impact of $\Delta$ and Sample Size Estimation}
\label{sub:kyberslash}
To study our proposed sample size estimation and the impact of different detection thresholds for $\Delta$, we test the reference implementation of Kyber, a post-quantum key encapsulation mechanism, for the Kyberslash 1 vulnerability~\cite{bernstein2025kyberslash}. Kyberslash 1 exploits runtime differences of secret-dependent division operations included when compiling the reference code optimized for code size on certain platforms. As Kyberslash provides a rather small side channel of \~20 cycles, we consider it a good example to conduct an empirical study on the statistical power we obtain when varying the parameters after estimating the required sample size. We provide more details on the collection of the measurements in \autoref{sec:real-world-eval-setup}.

\paragraph{Data Collection}
We begin our empirical study with a small sample of 300 measurements. Using our power analysis, we estimate the sample size needed to detect a leak of at least size~$\mu$ with detection rate~$p$, based on the observed variance and a set negligibility threshold~$\Delta$. Note that the actual leak of Kyberslash is fixed at around 20 cycles. By varying~$\mu$, we simulate a misestimation of the leak size and examine its effect on the required sample size and actual detection rate.
Similarly, we can adjust $\Delta$ to study the effect of lower and higher thresholds. Note that $\mu$ is solely a parameter for the statistical power estimation and not a parameter of our statistical test. As the side channel is fixed, the empirical detection rate ultimately only depends on the configured threshold ($\Delta$) and the estimated sample size ($n$). For each case, we evaluate the obtained sample size based on 1,000 data sets to study the detection rate empirically.

\paragraph{Empirical Results} 
\begin{table}
    \footnotesize
    \centering
    \caption{Results of our statistical power analysis based on an initial data set of 300 measurements for the Kyberslash leak. 
    On the right, we then present the empirical rejection rate (\textit{empirical p}) over 1,000 data sets with the estimated sample size for a fixed false positive rate of $\alpha = 0.1$.}
    \begin{tabular}{cccccc}
    \toprule
    \textbf{Case} & \textbf{$\mu$} & \textbf{$\Delta$} & \textbf{Targeted p} &
    \textbf{Estimated $n$} &  \textbf{Empirical p} \\
    \midrule
    1 & 15 & 5 & 0.9 & 1826 & 0.924 \\
    2 & 20 & 5 & 0.9 & 785 & 0.689 \\
    3 & 25 & 5 & 0.9 & 441 & 0.442 \\
    4 & 20 & 1 & 0.9 & 455 & 0.552 \\
    5 & 20 & 10 & 0.9 & 1683 & 0.841 \\
    6 & 20 & 5 & 0.75 & 359 & 0.377 \\
    7 & 20 & 5 & 0.99 & 793 & 0.707 \\
    8 & 50 & 40 & 0.9 & 1738 & 0.110 \\
    \bottomrule
\end{tabular}

    \label{tab:kyberEval-results}
\end{table}


We consider eight cases as shown in \autoref{tab:kyberEval-results}. Cases one to three use a fixed threshold~$\Delta$ of five cycles and a targeted detection rate~$p$ of 90\% but vary the expected size of the leak to be 15, 20, or 25 cycles, respectively. As we expect the real leak to be around 20 cycles based on the findings of~\cite{bernstein2025kyberslash}, our sample size estimation in case one should be too conservative as an assumed leak of 15 cycles is less pronounced than a leak of 20 cycles and thus requires a larger sample size for reliable detection. Indeed, we find that the estimated sample size of 1826 measurements results in a detection rate of 92.4\%, which is slightly above the targeted 90\%. Assuming a leak of 20 cycles in case 2,  the estimated sample size already drops to 785 measurements. However, with this sample size, we obtain an empirical detection rate of 68.9\% in contrast to the 90\% rate we targeted. This can be attributed to a higher variance contained in some of the 1,000 measurement sets we collected. Here, we emphasize again that the estimated sample size $n$ should be viewed as a guideline rather than an exact requirement. Finally, case three overestimates the leak to be 25 cycles, resulting in a proposed sample size of 441 cycles, which results in a detection rate of 44.2\%. Here, we expected a significantly lower detection rate as the $\mu$ value was not chosen appropriately. 

As illustrated by cases four and five, the threshold $\Delta$ also affects the estimated sample size. Generally, the closer the threshold is to the sought leak size $\mu$, the more measurements are necessary. Similarly, the targeted detection rate also affects the estimated sample size, as illustrated by cases six and seven. Note that while we can see the effect of the sample size on the empirical statistical power, they again remain well below the configured detection rate. A general takeaway is that collecting slightly more measurements is always a reasonable and safe approach.

As our last parameter example, we increase the threshold and expected leak size to $\Delta = 40$ and $\mu = 50$ in case eight. This simulates a scenario where the developer considers a leak of up to 40 cycles to be entirely negligible in their attacker model and is primarily looking to detect leaks of at least 50 cycles. As the Kyberslash leak lies below these thresholds, we expect to reach a detection rate close to our configured false positive rate, which we set to be 10\% in all tests. As $\Delta$ and $\mu$ are quite close again, we obtained a comparatively high sample size estimate of 1738, which yields an empirical detection rate of 11\%.

\subsection{Varying Measurement Scenarios}
\label{sub:webapp}
To study the outcomes of the sample size estimation for varying measurement scenarios, we implemented a web server that provides a timing leak during user authentication. This leak arises from the two-step authentication process: first, the server queries the database to check if the user is known and only then hashes and compares the password. A similar artificial leak used by Schinzel~\cite{mona12} to study tests for timing side channels.
To study the accuracy of the statistical power analysis, we complement it with varying measurement scenarios simulating a co-located, local LAN, and remote attacker introducing various levels of variance in the measurements. 
We provide more details on our implemented web server in \autoref{sec:real-world-eval-setup}. In our experiment, we now consider the view of an attacker who wants to know if a specific user is in the database of the server. The attacker uses a naive Python-based setup to measure server response times in nanoseconds. To estimate structural bias, they first collect 10,000 measurements using arbitrary usernames unknown to the server. They then estimate the side-channel timing gap by comparing responses for a known and an unknown username. Both values are obtained by the median of the absolute difference of the deciles. With both values in hand, the attacker can choose proper values in the statistical power analysis for the expected side-channel size and delta, such that the attacker can optimize a mass enumeration attack. 

\paragraph{Empirical Results}
\begin{table}
    \footnotesize
    \centering
    \caption{Results of our statistical power analysis based on an initial data set of 10,000 measurements for our web application. 
    On the right, we present the actual sample size required to detect the leak compared to the estimates sample size (estimated n) for a false positive rate of $\alpha$ = 0.1.}
    \begin{tabular}{clccccc}
        \toprule
        \textbf{Case} & \textbf{Description} & \textbf{$\mu$} & \textbf{$\Delta$} & \textbf{p} &
        \makecell[c]{\textbf{Estimated}\\ \textbf{$n$}} & \makecell[c]{\textbf{Test}\\ \textbf{Result}} \\
        \midrule
        1 & Local & 11,364 & 2,355 & 0.9 & 1,002  & \faStopwatch~(1,500) \\
        2 & LAN & 12,940 & 1,258 & 0.9  & 1,061  & \faStopwatch~(1,200) \\
        3 & WAN (local) & 38,479 & 4,975 & 0.9 & 1,846 & \faStopwatch~(7,000) \\
        4 & WAN (Inter.Cont.) & 44,667 & 52,773 & - & - & \faSlash \\
        \midrule
        \multicolumn{7}{c}{\faStopwatch~Difference (actual sample size) \hspace{5mm} \faSlash~No Difference}\\
        \bottomrule
    \end{tabular}
    \label{tab:webEval-results}
\end{table}
For the four cases, we obtained similar signal estimates $\mu$ for the Local and LAN setups, while the WAN setups showed varying signal strengths. We attribute this to increased variance in the WAN measurements, which reduces the reliability of the estimator for $\mu$ and $\Delta$. As a result, the statistical power analysis produced similar estimated sample sizes $n$ for the Local and LAN setups. These estimates slightly underestimated the number of measurements needed to detect a true positive. We believe this is due to a non-optimized measurement setup that led to a mild overestimation of $\mu$ and/or underestimation of $\Delta$. For example, in the Local setting, detection required 1,500 samples, which is far above the estimated 895.
The effect is even stronger visible in the first WAN case, where an estimated sample size of 1,846 was given, while in reality we need roughly 7,000 measurements. The reason for that can be explained by the poor estimation of $\mu$, $\Delta$ and/or unreliability of the measurement setup.

The results were even less reliable in the second WAN case. Due to the pronounced increase in variance, both $\mu$ and $\Delta$ were poorly estimated. For 10,000 samples, the estimate for $\Delta$ exceeded that of $\mu$, making statistical power analysis infeasible since it requires the signal $\mu$ to be greater than $\Delta$. However, if we reuse the Local parameters for $\mu$ and $\Delta$ in this fourth case, the estimated sample size jumps to $n = 396,562$. This sharp increase clearly illustrates the effect of high variance, as all other parameters remain the same.

Overall, it is visible that the measurement setup is a clear limitation of the procedure. The less reliable the measurements are, the less reliable the statistical guarantees. Consequently, some effort should also be put into optimizing the measurement setup. Possible options to do so involve isolating the measuring and measured process to separate cores (when measuring locally) and employing a proxy process to mitigate side affects between the test vector generation and the measurement collection. Alternatively, high precision hardware timestamping can be used, as done by Merget et al.~\cite{raccoon21}.
 


\subsection{Discussion on Sample Size Estimation}
In general, we find our statistical power analysis to provide a good first estimate for the sample size $n$. However, based on the empirical detection rates we obtained in Table \ref{tab:kyberEval-results} and the classifications in Table \ref{tab:webEval-results}, we stress that it may be beneficial to extend the estimated sample size. To this point, the largest sample size we report is based on the median of the variances over the decile differences. Alternatively, it would also be reasonable to report the minimum, the median, and the maximum for a better overview. On the other hand, presenting three sample sizes can be counterintuitive to end users. Ultimately, no single choice is \textit{always} best. Ideally, the estimation would always consider the variance of the quantile in which the signal of the leak manifests. However, this quantile is usually not known beforehand. Hence, we opted for the median of the quantile variances. It would also be possible to set $\mu$ slightly lower than the sought leak size to obtain a more conservative estimate. On top of that, it might also be reasonable to measure a larger initial data set, as this would capture deviations in the measurement setup more accurately, resulting in better estimations.

\section{Related Work}

\paragraph{Timing Attacks}
In 1996, Kocher first introduced the concept of timing attacks~\cite{C:Kocher96}.  In his work, he demonstrated how secret-dependent execution times in asymmetric cryptographic algorithms like RSA and DSS could be exploited. 
Subsequently, Tsunoo et al.~\cite{ches-2003-796} extended these attacks to symmetric cryptography and presented timing attacks on DES by exploiting cache timing.
Brumley and Boneh~\cite{brumley03} advanced this further by showing that timing attacks could be performed remotely by measuring variations in response times of TLS servers over a network. This discovery had significant implications for real-world security, leading to further attacks e.g. on TLS~\cite{lucky13,luckyMicroseconds15,pseudoSecure18}, WPA3~\cite{SP:VanRon20}, recent post-quantum cryptographic schemes~\cite{10.1007/978-3-030-56880-1_13,10.1007/978-3-030-38471-5_22}, and web applications\footnote{\scriptsize{\url{https://portswigger.net/research/listen-to-the-whispers-web-timing-attacks-that-actually-work}}}.
Kaufman et al.~\cite{10.1007/978-3-319-48965-0_36} further showed that vulnerabilities could persist even after compilation, highlighting the challenges of mitigating timing-based side channels.

\paragraph{Hardware Side Channels}
A lot of research on side channels focuses on hardware side channels, where the TVLA framework~\cite{tvla} is dominant in the literature. When analyzing hardware implementations, the tester can get closer to the source. The tester, therefore, can typically get better control over the system under test and control outside influences, which changes the 
distributions in the gathered data. Additionally, their measurements are often normally distributed, which justifies using tests like the t-test for the analysis. 
\paragraph{Analysis Techniques}
Analysis techniques for side channels can be roughly divided into four categories. Static analysis inspects the program code without running it. It usually requires users to annotate secret or sensitive values. This inspection can target the binary itself~\cite{USENIX:DFKMR13}, the LLVM IR~\cite{ctverif16}, or the source code\footnote{\scriptsize{\url{http://web.archive.org/web/20200810074547/http://trust-in-soft.com/tis-ct/}}}.
In contrast, dynamic analysis runs the annotated program with varying input values on a real system. It then observes the program flow to detect potential leaks. The main limitation is coverage. It can only guarantee the absence of leaks for the tested execution paths and input values. Examples of such tools include~\cite{ctgrind10, url:timecop, ctfuzz19, microwalk18, microwalkCi22, data18}.
Another approach is symbolic execution. It also executes the code but uses symbolic inputs instead of concrete ones. This technique allows to reason about the reachability and potential operands of each instruction. Symbolic techniques also typically require annotating secret or sensitive values. Examples of symbolic analysis tools include ~\cite{cached17, casym19, SP:DanBarRez20}.

\paragraph{Statistical Techniques}
Statistical analysis has always been important for evaluation timing side channels, especially outside academic research. An important step in this field came by Crosby et al.~\cite{crosby09} when investigating how timing vulnerabilities could be measured in remote scenarios. They proposed the box test, which was later implemented in \gls{mona}~\cite{mona12} and several tools presented at Blackhat~\cite{lawson10, timetrial14, morgan15}. Fu et al.~\cite{fu21} released a follow-up preprint of the \gls{dudect} paper. Their work also considers the Welch's t-test. In context of t-tests, Shelton at al.~\cite{rosita} also considered relevant hypotheses counterpart.
Fei et al.~\cite{statisticalModel14} introduced a theoretical framework using likelihood methods to analyze the theoretical properties of rather general side-channel attacks in more detail. 
Zhang et al.~\cite{zhangAes16} developed a theoretical model for cache-based side-chanel attacks also accounting for misclassifications.


\section{Conclusion}

In this work, we presented a new approach for statistical side-channel analysis that strives away from the classical hypothesis testing approach towards a relevant hypothesis approach. This approach gives users the ability to define which kind of side-channel sizes they care about and only find vulnerabilities that have \emph{at least} the given size with sound statistical guarantees. 
Additionally, our approach works with discrete and continuous data, making it applicable in even more scenarios. In general, our approach extends the set of statistical guarantees that can be provided to users when performing statistical analysis which we expect to assist statistical analysis to get broader adoption.

\paragraph{Future Work}
An important assumption that we are making in our algorithm is that the distributions from which we are drawing are \emph{strictly stationary} (see \autoref{Assumptions}). This means that the system under test should not change during our measurements \emph{independent} of our measurements. Just like all other approaches introduced so far, a strong violation of that assumptions can compromise the statistical guarantees. While our test can practically handle small changes, large changes will invalidate our statistical guarantees. Non-stationary measurements can appear in real measurements, when, for example, the network over which the measurements are being taken suddenly gets under heavy load, as this can drastically change the variance in the measurements. To account for such changes, a statistical algorithm could estimate when the distributions are changing and then use a testing procedure that accounts for these changes in behavior. Analyzing algorithms like this would further strengthen the robustness of the statistical analysis in practical relevant scenarios.

\bibliographystyle{plain}

\bibliography{main.bbl}

\begin{thebibliography}{10}

\bibitem{lucky13}
Nadhem~J. Al~Fardan and Kenneth~G. Paterson.
\newblock Lucky thirteen: Breaking the {TLS} and {DTLS} record protocols.
\newblock In {\em 2013 IEEE Symposium on Security and Privacy}.

\bibitem{luckyMicroseconds15}
Martin~R. Albrecht and Kenneth~G. Paterson.
\newblock Lucky microseconds: A timing attack on amazon's s2n implementation of {TLS}.
\newblock Cryptology ePrint Archive, Paper 2015/1129.

\bibitem{ctverif16}
Jos\'{e}~Bacelar Almeida, Manuel Barbosa, Gilles Barthe, Fran\c{c}ois Dupressoir, and Michael Emmi.
\newblock Verifying constant-time implementations.
\newblock In {\em Proceedings of the 25th USENIX Security Symposium}.

\bibitem{CCS:BBCLP14}
Gilles Barthe, Gustavo Betarte, Juan~Diego Campo, Carlos~Daniel Luna, and David Pichardie.
\newblock System-level non-interference for constant-time cryptography.
\newblock In Gail-Joon Ahn, Moti Yung, and Ninghui Li, editors, {\em ACM CCS 2014: 21st Conference on Computer and Communications Security}, pages 1267--1279. {ACM} Press, November 2014.

\bibitem{Bernstein2005CachetimingAO}
Daniel~J. Bernstein.
\newblock Cache-timing attacks on aes.
\newblock 2005.

\bibitem{bernstein2025kyberslash}
Daniel~J. Bernstein, Karthikeyan Bhargavan, Shivam Bhasin, Anupam Chattopadhyay, Tee~Kiah Chia, Matthias~J. Kannwischer, Franziskus Kiefer, Thales~B. Paiva, Prasanna Ravi, and Goutam Tamvada.
\newblock Kyberslash: Exploiting secret-dependent division timings in kyber implementations.
\newblock {\em IACR Transactions on Cryptographic Hardware and Embedded Systems}, 2025.
\newblock To appear.

\bibitem{C:Bleichenbacher98}
Daniel Bleichenbacher.
\newblock Chosen ciphertext attacks against protocols based on the {RSA} encryption standard {PKCS} \#1.
\newblock In Hugo Krawczyk, editor, {\em Advances in Cryptology -- {CRYPTO}'98}, volume 1462 of {\em Lecture Notes in Computer Science}, pages 1--12. Springer, Berlin, Heidelberg, August 1998.

\bibitem{bonferroni1936}
Carlo Bonferroni.
\newblock Teoria statistica delle classi e calcolo delle probabilita.
\newblock {\em Pubblicazioni del R istituto superiore di scienze economiche e commericiali di firenze}, 8:3--62, 1936.

\bibitem{bortzweb}
Andrew Bortz and Dan Boneh.
\newblock Exposing private information by timing web applications.
\newblock In {\em Proceedings of the 16th International Conference on World Wide Web}, WWW '07, page 621–628, New York, NY, USA, 2007. Association for Computing Machinery.

\bibitem{casym19}
Robert Brotzman, Shen Liu, Danfeng Zhang, Gang Tan, and Mahmut Kandemir.
\newblock {CaSym}: Cache aware symbolic execution for side channel detection and mitigation.
\newblock In {\em 2019 IEEE Symposium on Security and Privacy (SP)}.

\bibitem{stillPractical11}
Billy~Bob Brumley and Nicola Tuveri.
\newblock Remote timing attacks are still practical.
\newblock In Vijay Atluri and Claudia Diaz, editors, {\em Computer Security -- ESORICS 2011}.

\bibitem{brumley03}
David Brumley and Dan Boneh.
\newblock Remote timing attacks are practical.
\newblock In {\em 12th USENIX Security Symposium}, 2003.

\bibitem{CCS:CheFenDil17}
Jia Chen, Yu~Feng, and Isil Dillig.
\newblock Precise detection of side-channel vulnerabilities using quantitative cartesian hoare logic.
\newblock In Bhavani~M. Thuraisingham, David Evans, Tal Malkin, and Dongyan Xu, editors, {\em ACM CCS 2017: 24th Conference on Computer and Communications Security}, pages 875--890. {ACM} Press, October~/~November 2017.

\bibitem{crosby09}
Scott~A. Crosby, Dan~S. Wallach, and Rudolf~H. Riedi.
\newblock Opportunities and limits of remote timing attacks.
\newblock {\em ACM Trans. Inf. Syst. Secur.}, 2009.

\bibitem{SP:DanBarRez20}
Lesly-Ann Daniel, S{\'e}bastien Bardin, and Tamara Rezk.
\newblock Binsec/rel: Efficient relational symbolic execution for constant-time at binary-level.
\newblock In {\em 2020 {IEEE} Symposium on Security and Privacy}, pages 1021--1038. {IEEE} Computer Society Press, May 2020.

\bibitem{tvlapairedttest}
A.~Adam Ding, Cong Chen, and Thomas Eisenbarth.
\newblock Simpler, faster, and more robust t-test based leakage detection.
\newblock In Fran{\c{c}}ois-Xavier Standaert and Elisabeth Oswald, editors, {\em Constructive Side-Channel Analysis and Secure Design}, pages 163--183, Cham, 2016. Springer International Publishing.

\bibitem{USENIX:DFKMR13}
Goran Doychev, Dominik Feld, Boris K{\"o}pf, Laurent Mauborgne, and Jan Reineke.
\newblock {CacheAudit}: {A} tool for the static analysis of cache side channels.
\newblock In Samuel~T. King, editor, {\em USENIX Security 2013: 22nd {USENIX} Security Symposium}, pages 431--446. {USENIX} Association, August 2013.

\bibitem{USENIX:DMEMBS24}
Martin Dunsche, Marcel Maehren, Nurullah Erinola, Robert Merget, Nicolai Bissantz, Juraj Somorovsky, and J{\"o}rg Schwenk.
\newblock With great power come great side channels: Statistical timing side-channel analyses with bounded type-1 errors.
\newblock In Davide Balzarotti and Wenyuan Xu, editors, {\em USENIX Security 2024: 33rd {USENIX} Security Symposium}. {USENIX} Association, August 2024.

\bibitem{statisticalModel14}
Yunsi Fei, A.~Adam Ding, Jian Lao, and Liwei Zhang.
\newblock A statistics-based fundamental model for side-channel attack analysis.
\newblock Cryptology ePrint Archive, Paper 2014/152, 2014.

\bibitem{friedman-test}
Milton Friedman.
\newblock {A Comparison of Alternative Tests of Significance for the Problem of $m$ Rankings}.
\newblock {\em The Annals of Mathematical Statistics}, 11(1):86 -- 92, 1940.

\bibitem{fu21}
Xiaohan Fu, Shuyi Ni, Siyuan Yu, Zirui Wang, and Siwei Liu.
\newblock Constant-time analysis for well-known cryptography libraries.
\newblock Draft(Extended Abstract), 2021.

\bibitem{Qtools}
Marco Geraci.
\newblock Qtools: A collection of models and tools for quantile inference.
\newblock {\em The R Journal}, 8(2):117--138, 2016.

\bibitem{hiddencaches}
Matteo Golinelli and Bruno Crispo.
\newblock Hidden web caches discovery.
\newblock In {\em Proceedings of the 27th International Symposium on Research in Attacks, Intrusions and Defenses}, RAID '24, page 65–76, New York, NY, USA, 2024. Association for Computing Machinery.

\bibitem{10.1007/978-3-030-56880-1_13}
Qian Guo, Thomas Johansson, and Alexander Nilsson.
\newblock A key-recovery timing attack on post-quantum primitives using the fujisaki-okamoto transformation and its application on frodokem.
\newblock In {\em Advances in Cryptology – CRYPTO 2020: 40th Annual International Cryptology Conference, CRYPTO 2020, Santa Barbara, CA, USA, August 17–21, 2020, Proceedings, Part II}, page 359–386, Berlin, Heidelberg, 2020. Springer-Verlag.

\bibitem{lrv}
Sheila Görz and Alexander Dürre.
\newblock {\em robcp: Robust Change-Point Tests}, 2025.
\newblock R package version 0.3.8.

\bibitem{Hamilton1994}
James~D. Hamilton.
\newblock {\em Time Series Analysis}.
\newblock Princeton University Press, 1994.

\bibitem{bstar}
Tristen Hayfield and Jeffrey~S. Racine.
\newblock Nonparametric econometrics: The np package.
\newblock {\em Journal of Statistical Software}, 27(5):1--32, 2008.

\bibitem{ctfuzz19}
Shaobo He, Michael Emmi, and Gabriela~F. Ciocarlie.
\newblock ct-fuzz: Fuzzing for timing leaks.
\newblock {\em CoRR}, abs/1904.07280, 2019.

\bibitem{Holm1979}
Sture Holm.
\newblock A simple sequentially rejective multiple test procedure.
\newblock {\em Scandinavian Journal of Statistics}, 6(2):65--70, 1979.

\bibitem{SP:JFDSSB22}
Jan Jancar, Marcel Fourn{\'e}, Daniel {De Almeida Braga}, Mohamed Sabt, Peter Schwabe, Gilles Barthe, Pierre-Alain Fouque, and Yasemin Acar.
\newblock ``{T}hey're not that hard to mitigate'': What cryptographic library developers think about timing attacks.
\newblock In {\em 2022 {IEEE} Symposium on Security and Privacy}, pages 632--649. {IEEE} Computer Society Press, May 2022.

\bibitem{hardwarepktvla}
Aruna Jayasena, Emma Andrews, and Prabhat Mishra.
\newblock Tvla*: Test vector leakage assessment on hardware implementations of asymmetric cryptography algorithms.
\newblock {\em IEEE Transactions on Very Large Scale Integration (VLSI) Systems}, 31(9):1269--1279, 2023.

\bibitem{bootstrap_discrete}
Carsten Jentsch and Anne Leucht.
\newblock Bootstrapping sample quantiles of discrete data.
\newblock {\em Annals of the Institute of Statistical Mathematics}, 68(3):491--539, Jun 2016.

\bibitem{timeprivacy}
Zihao Jin, Ziqiao Kong, Shuo Chen, and Haixin Duan.
\newblock Timing-based browsing privacy vulnerabilities via site isolation.
\newblock In {\em 2022 IEEE Symposium on Security and Privacy (SP)}, pages 1525--1539, 2022.

\bibitem{ESORICS:Kario23}
Hubert Kario.
\newblock Everlasting {ROBOT}: The marvin attack.
\newblock In Gene Tsudik, Mauro Conti, Kaitai Liang, and Georgios Smaragdakis, editors, {\em ESORICS~2023: 28th European Symposium on Research in Computer Security, Part~III}, volume 14346 of {\em Lecture Notes in Computer Science}, pages 243--262. Springer, Cham, September 2023.

\bibitem{kario23}
Hubert Kario.
\newblock Out of the box testing.
\newblock {\em {IACR} Cryptol. ePrint Arch.}, page 1441, 2023.

\bibitem{10.1007/978-3-319-48965-0_36}
Thierry Kaufmann, Herv{\'e} Pelletier, Serge Vaudenay, and Karine Villegas.
\newblock When constant-time source yields variable-time binary: Exploiting curve25519-donna built with msvc 2015.
\newblock In Sara Foresti and Giuseppe Persiano, editors, {\em Cryptology and Network Security}, pages 573--582, Cham, 2016. Springer International Publishing.

\bibitem{SP:KHFGGH19}
Paul Kocher, Jann Horn, Anders Fogh, Daniel Genkin, Daniel Gruss, Werner Haas, Mike Hamburg, Moritz Lipp, Stefan Mangard, Thomas Prescher, Michael Schwarz, and Yuval Yarom.
\newblock Spectre attacks: Exploiting speculative execution.
\newblock In {\em 2019 {IEEE} Symposium on Security and Privacy}, pages 1--19. {IEEE} Computer Society Press, May 2019.

\bibitem{C:Kocher96}
Paul~C. Kocher.
\newblock Timing attacks on implementations of {Diffie}-{Hellman}, {RSA}, {DSS}, and other systems.
\newblock In Neal Koblitz, editor, {\em Advances in Cryptology -- {CRYPTO}'96}, volume 1109 of {\em Lecture Notes in Computer Science}, pages 104--113. Springer, Berlin, Heidelberg, August 1996.

\bibitem{ctgrind10}
Adam Langley.
\newblock ctgrind: Checking that functions are constant time with valgrind.
\newblock \url{https://github.com/agl/ctgrind}.

\bibitem{lawson10}
Nate Lawson and Taylor Nelson.
\newblock Exploiting timing attacks in widespread systems, Blackhat 2010.

\bibitem{tvla}
Tal Lev-Ami and Mooly Sagiv.
\newblock Tvla: A system for implementing static analyses.
\newblock In Jens Palsberg, editor, {\em Static Analysis}, pages 280--301, Berlin, Heidelberg, 2000. Springer Berlin Heidelberg.

\bibitem{USENIX:LSGPHF18}
Moritz Lipp, Michael Schwarz, Daniel Gruss, Thomas Prescher, Werner Haas, Anders Fogh, Jann Horn, Stefan Mangard, Paul Kocher, Daniel Genkin, Yuval Yarom, and Mike Hamburg.
\newblock Meltdown: Reading kernel memory from user space.
\newblock In William Enck and Adrienne~Porter Felt, editors, {\em USENIX Security 2018: 27th {USENIX} Security Symposium}, pages 973--990. {USENIX} Association, August 2018.

\bibitem{ma2011asymptotic}
Yanyuan Ma, Marc~G Genton, and Emanuel Parzen.
\newblock Asymptotic properties of sample quantiles of discrete distributions.
\newblock {\em Annals of the Institute of Statistical Mathematics}, 63:227--243, 2011.

\bibitem{timetrial14}
Daniel~A. Mayer and Joel Sandin.
\newblock Time trial, racing towards practical remote timing attacks, Blackhat 2014.

\bibitem{raccoon21}
Robert Merget, Marcus Brinkmann, Nimrod Aviram, Juraj Somorovsky, Johannes Mittmann, and J{\"o}rg Schwenk.
\newblock Raccoon attack: Finding and exploiting most-significant-bit-oracles in {TLS-DH(E)}.
\newblock In {\em Proceedings of the 30th USENIX Security Symposium}, 2021.

\bibitem{meyer14}
Christopher Meyer, Juraj Somorovsky, Eugen Weiss, J{\"o}rg Schwenk, Sebastian Schinzel, and Erik Tews.
\newblock Revisiting {SSL$/$TLS} implementations: New {Bleichenbacher} side channels and attacks.
\newblock In {\em 23rd USENIX Security Symposium}, 2014.

\bibitem{chi2tvla}
Amir Moradi, Bastian Richter, Tobias Schneider, and François-Xavier Standaert.
\newblock Leakage detection with the x2-test.
\newblock {\em IACR Transactions on Cryptographic Hardware and Embedded Systems}, pages 209--237, 02 2018.

\bibitem{morgan15}
Timothy~D. Morgan and Jason~W. Morgan.
\newblock Web timing attacks made practical, Blackhat 2015.

\bibitem{Noble2009}
William~Stafford Noble.
\newblock How does multiple testing correction work?
\newblock {\em Nature Biotechnology}, 27(12):1135--1137, 2009.

\bibitem{10.1007/978-3-030-38471-5_22}
Thales~Bandiera Paiva and Routo Terada.
\newblock A timing attack on the hqc encryption\&nbsp;scheme.
\newblock In {\em Selected Areas in Cryptography – SAC 2019: 26th International Conference, Waterloo, ON, Canada, August 12–16, 2019, Revised Selected Papers}, page 551–573, Berlin, Heidelberg, 2019. Springer-Verlag.

\bibitem{Politis31122004}
Dimitris~N. Politis and Halbert~White and.
\newblock Automatic block-length selection for the dependent bootstrap.
\newblock {\em Econometric Reviews}, 23(1):53--70, 2004.

\bibitem{Ragab2021CrossTalkSD}
Hany Ragab, Alyssa Milburn, Kaveh Razavi, Herbert Bos, and Cristiano Giuffrida.
\newblock Crosstalk: Speculative data leaks across cores are real.
\newblock {\em 2021 IEEE Symposium on Security and Privacy (SP)}, pages 1852--1867, 2021.

\bibitem{dudect16}
Oscar Reparaz, Josep Balasch, and Ingrid Verbauwhede.
\newblock Dude, is my code constant time?
\newblock Cryptology ePrint Archive, Paper 2016/1123, 2016.

\bibitem{pseudoSecure18}
Eyal Ronen, Kenneth~G. Paterson, and Adi Shamir.
\newblock Pseudo constant time implementations of {TLS} are only pseudo secure.
\newblock In {\em Proceedings of the 2018 ACM SIGSAC Conference on Computer and Communications Security}.

\bibitem{Sachs.1984}
Lothar. Sachs.
\newblock {\em Applied Statistics: A Handbook of Techniques}.
\newblock 2 edition, 1984.

\bibitem{rsacrttiming}
Werner Schindler.
\newblock A timing attack against rsa with the chinese remainder theorem.
\newblock volume 1965, pages 109--124, 01 2000.

\bibitem{mona12}
Sebastian Schinzel.
\newblock {\em Unintentional and Hidden Information Leaks in Networked Software Applications}.
\newblock doctoralthesis, Friedrich-Alexander-Universit{\"a}t Erlangen-N{\"u}rnberg, 2012.

\bibitem{rosita}
Madura~A. Shelton, Łukasz Chmielewski, Niels Samwel, Markus Wagner, Lejla Batina, and Yuval Yarom.
\newblock Rosita++: Automatic higher-order leakage elimination from cryptographic code.
\newblock Cryptology {ePrint} Archive, Paper 2021/1181, 2021.

\bibitem{bonferroni}
SpringerLink.
\newblock Bonferroni correction, 27.04.2023.

\bibitem{tvlanistlwc}
Thomas Steinbauer, Rishub Nagpal, Robert Primas, and Stefan Mangard.
\newblock {TVLA} on selected {NIST LWC} finalists.
\newblock {\em Cryptology ePrint Archive}, 2022(887), August 2022.
\newblock Cryptology ePrint Archive, Report 2022/887.

\bibitem{url:timecop}
{TIMECOP}.
\newblock \url{https://www.post-apocalyptic-crypto.org/timecop/}, Accessed: 05.05.2023.

\bibitem{ches-2003-796}
Yukiyasu Tsunoo, Teruo Saito, Tomoyasu Suzaki, Maki Shigeri, and Hiroshi Miyauchi.
\newblock Cryptanalysis of des implemented on computers with cache.
\newblock In {\em Cryptographic Hardware and Embedded Systems - CHES 2003, 5th International Workshop, Cologne, Germany, September 8-10, 2003, Proceedings}, volume 2779 of {\em Lecture Notes in Computer Science}, pages 62--76. Springer, 2003.

\bibitem{Tukey1991ThePO}
John~W. Tukey.
\newblock The philosophy of multiple comparisons.
\newblock {\em Statistical Science}, 6:100--116, 1991.

\bibitem{tvlapublickey}
Michael Tunstall and Gilbert Goodwill.
\newblock Applying tvla to public key cryptographic algorithms.
\newblock {\em IACR Cryptol. ePrint Arch.}, 2016:513, 2016.

\bibitem{vaart1998}
A.~W. van~der Vaart.
\newblock {\em Asymptotic Statistics}.
\newblock Cambridge Series in Statistical and Probabilistic Mathematics. Cambridge University Press, 1998.

\bibitem{timewilltell}
Vik Vanderlinden, Tom~Van Goethem, and Mathy Vanhoef.
\newblock Time will tell: Exploiting timing leaks using http response headers.
\newblock In {\em Computer Security – ESORICS 2023: 28th European Symposium on Research in Computer Security, The Hague, The Netherlands, September 25–29, 2023, Proceedings, Part II}, page 3–22, Berlin, Heidelberg, 2024. Springer-Verlag.

\bibitem{SP:VanRon20}
Mathy Vanhoef and Eyal Ronen.
\newblock Dragonblood: Analyzing the dragonfly handshake of {WPA3} and {EAP}-pwd.
\newblock In {\em 2020 {IEEE} Symposium on Security and Privacy}, pages 517--533. {IEEE} Computer Society Press, May 2020.

\bibitem{math-statistics-book-sign-test}
Dennis~D. Wackerly, William Mendenhall, and Richard~L. Scheaffer.
\newblock {\em Mathematical statistics with applications}.
\newblock Duxbury, 7th edition, 2007.

\bibitem{cached17}
Shuai Wang, Pei Wang, Xiao Liu, Danfeng Zhang, and Dinghao Wu.
\newblock {CacheD}: Identifying cache-based timing channels in production software.
\newblock In {\em Proceedings of the 26th USENIX Security Symposium}, 2017.

\bibitem{USENIX:WPHSFK22}
Yingchen Wang, Riccardo Paccagnella, Elizabeth~Tang He, Hovav Shacham, Christopher~W. Fletcher, and David Kohlbrenner.
\newblock Hertzbleed: Turning power side-channel attacks into remote timing attacks on {x86}.
\newblock In Kevin R.~B. Butler and Kurt Thomas, editors, {\em USENIX Security 2022: 31st {USENIX} Security Symposium}, pages 679--697. {USENIX} Association, August 2022.

\bibitem{USENIX:WSBS20}
Samuel Weiser, David Schrammel, Lukas Bodner, and Raphael Spreitzer.
\newblock Big numbers - big troubles: Systematically analyzing nonce leakage in ({EC}){DSA} implementations.
\newblock In Srdjan Capkun and Franziska Roesner, editors, {\em USENIX Security 2020: 29th {USENIX} Security Symposium}, pages 1767--1784. {USENIX} Association, August 2020.

\bibitem{data18}
Samuel Weiser, Andreas Zankl, Raphael Spreitzer, Katja Miller, Stefan Mangard, and Georg Sigl.
\newblock {DATA}--differential address trace analysis: Finding address-based side-channels in binaries.
\newblock In {\em Proceedings of the 27th USENIX Security Symposium}, 2018.

\bibitem{wellek2010}
Stefan Wellek.
\newblock {\em Testing Statistical Hypotheses of Equivalence and Noninferiority}.
\newblock Chapman and Hall/CRC, 2nd edition, 2010.

\bibitem{microwalk18}
Jan Wichelmann, Ahmad Moghimi, Thomas Eisenbarth, and Berk Sunar.
\newblock {MicroWalk}: {A} framework for finding side channels in binaries.
\newblock {\em CoRR}, abs/1808.05575, 2018.

\bibitem{microwalkCi22}
Jan Wichelmann, Florian Sieck, Anna Pätschke, and Thomas Eisenbarth.
\newblock Microwalk-{CI}.
\newblock In {\em Proceedings of the 2022 {ACM} {SIGSAC} Conference on Computer and Communications Security}.

\bibitem{wilcoxon-signed-rank}
Frank Wilcoxon.
\newblock Individual comparisons by ranking methods.
\newblock {\em Biometrics Bulletin}, 1(6):80--83, 1945.

\bibitem{ecdsaPractical15}
David Wong.
\newblock Timing and lattice attacks on a remote {ECDSA OpenSSL} server: How practical are they really?
\newblock Cryptology ePrint Archive, Paper 2015/839, 2015.

\bibitem{cachebleed17}
Yuval Yarom, Daniel Genkin, and Nadia Heninger.
\newblock {CacheBleed}: a timing attack on {OpenSSL} constant-time {RSA}.
\newblock {\em Journal of Cryptographic Engineering}, 7, 06 2017.

\bibitem{zhangAes16}
Liwei Zhang, A~Adam Ding, Yunsi Fei, and Zhen~Hang Jiang.
\newblock Statistical analysis for access-driven cache attacks against {AES}.
\newblock {\em Cryptology ePrint Archive}, 2016.

\bibitem{Sidak1967}
Zbyněk Šidák and.
\newblock Rectangular confidence regions for the means of multivariate normal distributions.
\newblock {\em Journal of the American Statistical Association}, 62(318):626--633, 1967.

\end{thebibliography}

\appendix
\section{Generating Dependent Data}\label{app:gen_data}
Since one major contribution of our approach is modeling dependency between observations and coordinates, we consider a very basic model that generates dependent observations, namely the autoregressive model of order one $AR(1)$ (cf. \cite{Hamilton1994} chapter 1). It is a fundamental time series model that expresses a variable $Y_t$ as a function of its past values (modeling the dependence) plus an additional noise term:
\begin{equation}\label{eq:ar_process}
    Y_n = \phi Y_{n-1} + \epsilon_n, \quad \epsilon_n \sim \mathcal{N}(0, \sigma^2)~.
\end{equation}
To determine the strength of the dependence, we chose $\phi$ between $-1$ and $1$, where values closer to $\pm 1$ correspond to stronger dependence between observations. The case $\phi=0$ corresponds to the classical normally distributed case with independent observations ($Y_n$ does not depend on $Y_{n-1}$ anymore). Since we consider two sample problems, we will consider observations of two $AR(1)$ processes of size $n$, say $x_1,\hdots, x_n$ and $y_1,\hdots, y_n$, where we shift all $y$ by $\mu$.\\
\textbf{Parameter settings:} 
Throughout the simulations, we let the dependence parameter be $\phi$ vary between $-0.9$ and $0.9$ and the signal $\mu$ between $0$ and $1$. For the Bootstrap procedures, we chose $B=1000$ throughout the simulations. 


\section{Illustration t-test}
\label{app:example_t_test}

\paragraph{Misleading Two Step Procedure}
Subsequently, we illustrate two distinct facts concerning relevant hypotheses. For that purpose, consider the well-known one-sample t-test with the following pair of hypotheses:
\begin{align}\label{eq:classical_t-test} 
H_0: \mu = 0 \quad \text{vs.} \quad H_1: \mu \neq 0~. 
\end{align}
For normally distributed data, the t-test is known to yield reliable results. As already indicated in Section \ref{subsec:advantages_relevant_hypothesis}, we only want to detect relevant changes of size $\Delta$ or larger. One might therefore consider testing the following hypotheses-pair instead: 
\begin{align}\label{eq:rel_t-test} 
H_0: |\mu| \leq \Delta \quad \text{vs.} \quad H_1: |\mu| > \Delta~.
\end{align}
To test the hypotheses-pair in \eqref{eq:rel_t-test} with a prescribed type-1 error $\alpha$, instead of adapting our approach in Section \ref{sec:hypothesis_test} to the t-test, it might also seem natural to consider a two-step procedure: First, perform the classical t-test and second reject only if $\bar x$ also exceeds~$\Delta$. However, already in this simple example, one can see that this can yield an inflated $\alpha$. In fact, if the true $\mu=\Delta$, we get by default $\alpha=0.5$, since the distribution of the empirical mean is still symmetric around $\mu$. Or in other words, in $50\%$ of the times it is greater than $\Delta$ and in $50\%$ it is smaller, yielding to a rejection rate of $0.5$, which is way higher than $\alpha$ set in the t-test. Simulating that two-step procedure also underpins that, yielding an empirical rejection rate of $\hat\alpha\approx 0.5$. The closer $\mu$ gets to zero, the less obscure this two-step procedure will be. \\ In this very simple example, it is not difficult to correctly reject. To properly reject the null hypothesis in \eqref{eq:rel_t-test}, the following rejection rule yields a level $\alpha$ test: reject $H_0$ if and only if $$\frac{|\bar x|-\Delta}{\sigma_{\bar x}}>c_{1-\alpha}~,$$
where $c_{1-\alpha}$ is the $1-\alpha$ quantile of the standard normal distribution. Note that only simplicity we assumed to know the standard deviation of $\bar x $. For more advanced hypotheses, like in \eqref{eq:null_hyp}, the procedure is way more involved and simple adaptations as above are not possible.
\paragraph{Reversing Hypotheses}
Considering the hypotheses-pair in \eqref{eq:rel_t-test}, we have already pointed out that it might also be reasonable to control the type-2 error $\beta$. For that purpose, we usually reverse hypotheses and therefore $\beta$ will be the type-1 error with respect to the following hypotheses-pair:
\begin{align*}
   H_0: |\mu| \geq \Delta \quad \text{vs.} \quad H_1: |\mu| < \Delta~. 
\end{align*}
This is indeed possible and yields no further complications. The rejection rule is only adapted by the following: we reject $H_0$ if and only if
$$\frac{|\bar x|-\Delta}{\sigma_{\bar x}}<c_{\alpha}~.$$
One question that might arise is whether reversing the hypotheses is also possible in the classical setting of \eqref{eq:classical_t-test}. The clear answer is no. If we were to set $H_0:\mu\neq 0$, rejecting $H_0$ would depend on the true (but unknown) value of $\mu$. To put it intuitively: imagine being in court and trying to prove someone is not guilty as the null. Then there are countless ways a person could be innocent, but only one specific crime they are accused of. For each of the ways a person is innocent, you would in principle have to define a rejection rule. Translating this to the statistical setting: for each possible $\mu$, the distribution of the test statistic is different (as the $\bar x$ follows a different distribution); hence the rejection rule is different for a prescribed $\alpha$.

\section{Additional Definitions and Algorithms}\label{app:add_def_alg}
For a discrete random variable $X$ taking values in $x_1,\hdots, x_d$ with probabilities $p_1,\hdots, p_d$, we define $F_{mid}(x)=F(x)-0.5p(x)$. Then, if $\pi_k:=\sum_{i=1}^{k-1} p_i+p_k/2$, the mid-distribution quantile function can be defined by
$$F_{mid}^{-1}(p):=\begin{cases}
    x_1,\quad & \text{if } p<p_1/2,\\
    x_k,\quad & \text{if } p=p_k, k=1,\hdots, d\\
    \lambda x_k+(1-\lambda)x_{k+1},\quad & \text{if }p=\lambda \pi_k+(1-\lambda)\pi_{k+1},\\
    x_d, \quad & \text{if } p>\pi_d
\end{cases}~,$$ 
where $0<\lambda<1$. As already mentioned in the previous section, the standard n-out-of-n bootstrap fails. We, therefore, adopt an m-out-of- n bootstrap procedure proposed in Jentsch and Leucht~\cite{bootstrap_discrete}.
\begin{algorithm}
    \small
    \noindent\textbf{function} \textsc{Bootstrap}($x$, $y$, $\alpha$, $\Delta$,$K_{sub}^{max}$, $B$, $m$)
    \begin{algorithmic}[1]
    \REQUIRE $x$, $y$, $\alpha$, $\Delta$, $K_{sub}^{max}$, $B$, $m$
    \ENSURE $\hat Q^*$

    \FOR{$i = 1, \hdots, B$}
        \STATE Set $m_1 = n^{2/3}$.
        \STATE Sample $I \subset \{1, \hdots, n - m + 1\}$ with $|I| = \lceil m_1/m \rceil$.
        \STATE Set $x^* = x[I], y^* = y[I]$, where each index $i \in I$ includes all elements $x_i, \hdots, x_{i+m-1}$.
        \STATE Compute bootstrap test-statistic:
        \[
        \hat Q^{i,*} = \sqrt{m_1} \left( \left| \hat q_k^{X,*} - \hat q_k^{Y,*} \right| - \left| \hat q_k^X - \hat q_k^Y \right| \right)
        \].
    \ENDFOR

    \STATE \textbf{return} $\hat Q^* := (\hat Q^{1,*}, \hdots, \hat Q^{B,*})$
    \end{algorithmic}
    \caption{Bootstrap Maximum for discrete data}
    \label{algorithm:bootstrap_max_discrete}
\end{algorithm}

\section{Proofs}
\begin{proof}{Proof of Theorem \autoref{thm:asymptotic}}\\
Assume that we are under the null hypothesis. Therefore, we can decompose $\hat Q$ by
\begin{align}\label{p1}
\hat Q&=\sqrt{n}\max_{k \in K}(|\hat q_k^X-\hat q_k^Y|-\Delta)\\
&= \max\Big\{\sqrt{n}\max_{k \in K,|\hat q_k^X-\hat q_k^Y|=\Delta}(|\hat q_k^X-\hat q_k^Y|-\Delta),\nonumber\\&\quad\sqrt{n}\max_{k \in K,|\hat q_k^X-\hat q_k^Y|<\Delta}(|\hat q_k^X-\hat q_k^Y|-\Delta)\Big\}\nonumber
\end{align}
and note that by Lemma 21.4 from \cite{vaart1998} combined with the asymptotic normality of the empirical process for $m$-dependent data the second quantity in the outer maximum diverges to $-\infty$. The first quantity converges in distribution against the limit 
\begin{align}
\label{p2}
\max_{k \in K,|q_k^X-q_k^Y|=\Delta}\text{sign}(q_k^X-q_K^Y)\left(\frac{\mathbb{G}_1(k)}{f_X(q_k^X)}-\frac{\mathbb{G}_2(k)}{f_Y(q_k^Y)}\right)~,
\end{align} due to the weak convergence of the quantile processes combined with the continuous mapping theorem. As the first quantity is thus also tight, it dominates the maximum. We hence obtain that, under $H_0$, the test statistic is given by
\begin{align}
\label{p6}
    \hat Q=\sqrt{n}\max_{k \in K,|\hat q_k^X-\hat q_k^Y|=\Delta}(|\hat q_k^X-\hat q_k^Y|-\Delta)
\end{align}
with high probability, this yields the desired result by \eqref{p2}.

\end{proof}
\begin{proof}{Proof of Theorem \autoref{thm:bootstraplevel}}\\
By Theorem \ref{thm:asymptotic} we only need to establish that the bootstrap process $\hat Q^*$ converges to the distribution given in equation \eqref{p2} in probability given the original data. 
To that end, we first note that Lemma 21.4 as well as Theorems 23.7 and 23.9  from \cite{vaart1998} yield that
\begin{align}
\label{p3}
    &\Big\{\sqrt{n}\left(\hat q^{X,*}_k-\hat q^{Y,*}_k-(\hat q_k^X-\hat q_k^Y)\right)\Big\}_{k \in K} \\&\qquad\overset{d}{\rightarrow}\Big\{\frac{\mathbb{G}_1(k)}{f_X(q_k^X)}-\frac{\mathbb{G}_2(k)}{f_Y(q_k^Y)}\Big\}_{k \in K}\nonumber
\end{align} in probability given $\mathbb{Y}:=\{X_1,...,X_n,Y_1,...,Y_n\}$.  This also yields
\begin{align}
\label{p4}
    &\max_{k \in K^*}\sqrt{n}\left(\hat q^{X,*}_k-\hat q^{Y,*}_k-(\hat q_k^X-\hat q_k^Y)\right)-\\
    &\qquad \max_{k \in K,|q^X_k-q^Y_k|=\Delta}\sqrt{n}\left(\hat q^{X,*}_k-\hat q^{Y,*}_k-(\hat q_k^X-\hat q_k^Y)\right)=o_\mathbb{P}(1)\nonumber
\end{align}
as the index sets are the same with high probability. This follows because 
we have by asymptotic normality that
\begin{align*}
&\mathbb{P}\left(\sup_{k \in K: |q^X_k-q^Y_k|=\Delta}(|\hat q^X_k-\hat q^Y_k|-\Delta) \in (-\sqrt{\log(n)/n)},\sqrt{\log(n)/n})\right)\\
&=\mathbb{P}\left(\sqrt{n}\sup_{k \in K: |q^X_k-q^Y_k|=\Delta}(|\hat q^X_k-\hat q^Y_k|-\Delta) \in (-\sqrt{\log(n))},\sqrt{\log(n)})\right)\\&\to 1~,
\end{align*}
so that $k \in K^*$ with probability tending to 1. For those $k$ with $|q^X_k-q^Y_k|\neq \Delta$ a similar argument shows that the probability of $k \in K^*$ goes to 0. Taking subsequences, we can turn \eqref{p3} into an almost sure statement, combining this with the continuous mapping theorem and equation \eqref{p4} yields the desired convergence along the subsequence. As we can do this for any subsequence, we obtain the result.

\end{proof}
\begin{proof}{Proof of Theorem \autoref{thm:bootstrappower}}\\
For the first statement of the theorem first note that $\hat Q$ diverges to $+\infty$, to be precise we have for any index $k_0 \in K$
\begin{align}
\label{p5}
    \hat Q&=\sqrt{n}\max_{k \in K}(|\hat q_k^X-\hat q_k^Y|-\Delta)\\
    &\geq \sqrt{n}(|\hat q_{k_0}^X-\hat q_{k_0}^Y|-|q_{k_0}^X-q_{k_0}^Y|)+\sqrt{n}(|q_{k_0}^X-q_{k_0}^Y|-\Delta)~.
\end{align}
The first summand in the last line is tight (uniformly in $k_0$) by Corollary 21.5 from \cite{vaart1998} and the continuous mapping theorem and the second summand diverges for at least one index $k_0 \in K$. Combining this with the fact that $c^*_{1-\alpha}$ is bounded in probability (see the proof of theorem \ref{thm:bootstraplevel}) yields the desired result.\\

For the second statement note that when $k_0$ denotes the unique index with $|q^X_{k_0}-q^Y_{k_0}|=\Delta$ equation \eqref{p5} becomes an equality with high probability by arguments similar to those that lead to equation \eqref{p6}. Consequently, we have
\begin{align}
    &\mathbb{P}(\hat Q > q_{1-\alpha})=o(1)+\\
    &\mathbb{P}(\sqrt{n}(|\hat q_{k_0}^X-\hat q_{k_0}^Y|-|q_{k_0}^X-q_{k_0}^Y|)>q_{1-\alpha}-\sqrt{n}(|q_{k_0}^X-q_{k_0}^Y|-\Delta))~.
\end{align}
We then obtain the desired result by noting that $(\sqrt{n}(|\hat q_{k_0}^X-\hat q_{k_0}^Y|-|q_{k_0}^X-q_{k_0}^Y|)$ converges against $Z$ by  the same arguments that yield \eqref{p2}.
\end{proof}

\section{Additional Details of the Real World Evaluation}
\label{sec:real-world-eval-setup}

\paragraph{Kyberslash}
When compiling the reference Kyber implementation released before the discovery of Kyberslash~\cite{bernstein2025kyberslash}  for minimal code size (\texttt{-Os}), an operand-dependent division operation is used on certain platforms. As one of these operands is a coefficient of the secret key used for decryption, this division operation poses a side channel that can be used to reconstruct the private key entirely~\cite{bernstein2025kyberslash}. For our data set, we collected measurements on a Raspberry Pi 2B that features an affected ARM CPU. Specifically, we adapted the demo script\footnote{\url{https://kyberslash.cr.yp.to/demos.html}} provided by the Kyberslash authors to measure two vectors for a fixed secret key: one expected to yield a slow division, and one expected to yield a fast division resulting in a difference of 20 cycles. In total, we performed 1,000 iterations each collecting 5,0000 measurements for each of the two vectors. For our comparison of test results with varying sample sizes, we then created subsets of these measurements.

\paragraph{Web Application}
We developed and evaluated a small Flask application written in Python, which uses SQLite as its database and SHA-256 for its hashing operations. All functionality was implemented using the official Python modules.

For each measurement scenario, we collected two sets of measurements. In the first set, we measured the server's response time when the request included an existing username and incorrect password. In the second set, we measured the response time when the request included a non-existent username and an incorrect password. We expect the response times in the first set to be longer because the existing username triggers additional internal processing (a hash operation and a comparison).

To collect these measurements, we deployed the web application on a server with an AMD EPYC 7763 CPU, 2 TB RAM, and Ubuntu 22.04.4 LTS.
For the LAN measurements, we used a client with the same hardware and operating system. 
For the WAN (local) measurements, we used a client with an AMD Ryzen 7 PRO 5850U, 48 GB of RAM, and Ubuntu 24.04.2 LTS.
For the WAN (Inter.Cont.) measurements, we used a client with an Intel Core i9-14900K,  64GB of RAM, and Ubuntu 24.04 LTS.

\end{document}